\documentclass[useAMS,usenatbib,babel]{mn2e}

\usepackage[english,english]{babel}
\usepackage{amsmath}
\usepackage{amssymb,amsfonts,textcomp}
\usepackage{array}
\usepackage{supertabular}
\usepackage{hhline}
\usepackage{hyperref}
\usepackage[usenames]{color}
\hypersetup{dvips, colorlinks=true, linkcolor=black, citecolor=black, filecolor=black, urlcolor=black}
\usepackage[dvips]{graphicx}

\def\gtrsim{\lower.5ex\hbox{$\; \buildrel > \over \sim \;$}}
\usepackage{graphicx}

\newcommand{\hagn}{\mbox{{\sc \small Horizon-AGN\,\,}}}

\definecolor{grey}{rgb}{0.75,0.75,0.75}
\definecolor{Orange}{rgb}{1.0,0.5,0.15}
\definecolor{brown}{rgb}{0.7,0.25,0.0}
\definecolor{pink}{rgb}{1.0,0.5,0.5}
\definecolor{darkerred}{rgb}{0.8,0,0}
\definecolor{darkerblue}{rgb}{0,0,0.8}
\definecolor{Blue}{rgb}{0,0.08,0.65}
\definecolor{Red}{rgb}{0.65,0.08,0.05}
\definecolor{Green}{rgb}{0.15,0.45,0.25}

\begin{document}

\author[C. Welker et al.]{
\parbox[t]{\textwidth}{C. Welker$^{1,2}$\thanks{E-mail: welker@iap.fr}, Y. Dubois$^{1}$,  J.~Devriendt$^{2,3}$, C. Pichon$^{1,4}$, S. Kaviraj$^{5}$
and S. Peirani$^{1}$}
\vspace*{6pt} \\ 
$^{1}$ CNRS and UPMC Univ. Paris 06, UMR 7095, Institut d'Astrophysique de Paris, 98 bis Boulevard Arago, F-75014 Paris, France\\
$^{2}$ Sub-department of Astrophysics, University of Oxford, Keble Road, Oxford OX1 3RH\\
$^{3}$ Observatoire de Lyon, UMR 5574, 9 avenue Charles Andr\'e, Saint Genis Laval 69561, France\\
$^{4}$ {Institute of Astronomy, University of Cambridge, Madingley Road, Cambridge, CB3 0HA}\\
$^{5}$ Centre for Astrophysics Research, University of Hertfordshire, College Lane, Hatfield, Herts, AL10 9AB\\
}
\date{Accepted . Received ; in original form }

\title[The rise and fall of $z>1$ disk galaxies]{The rise and fall of stellar disks across the peak of cosmic star formation history: mergers versus smooth accretion}

\maketitle

\begin{abstract}
{Building galaxy merger trees from a state-of-the-art cosmological hydrodynamics simulation, \hagn\!\!, we perform a statistical study of how mergers and smooth accretion drive galaxy morphologic properties above $z>1$.
More specifically, we investigate how stellar densities, effective radii and shape parameters derived from the inertia tensor depend on mergers of different mass ratios. 
We find strong evidence that smooth accretion tends to flatten small galaxies over cosmic time, leading to the formation of disks.
On the other hand, mergers, and not only the major ones, exhibit a propensity to puff up and destroy stellar disks, confirming the origin of elliptical galaxies. 
We also find that elliptical galaxies are more susceptible to grow in size through mergers than disc galaxies with a size-mass evolution $r_{0.5}\propto M_{\rm s}^{1.2}$ instead of $r_{0.5}\propto M_{\rm s}^{-0.5}-M^{0.5}$ depending on the merger mass ratio.
The gas content drive the size-mass evolution due to merger with a faster size growth for gas-poor galaxies $r_{0.5}\propto M_{\rm s}^{2}$ than for gas-rich galaxies $r_{0.5}\propto M_{\rm s}$.
}
\end{abstract}

\begin{keywords}
galaxies: formation ---
galaxies: evolution ---
galaxies: interactions ---
galaxies: kinematics and dynamics ---
methods: numerical
\end{keywords}

\section{Introduction}

While our present understanding of galaxy evolution derives mainly
from the nearby ($z < 1$) Universe, the bulk of today's stellar mass
formed around the broad peak of cosmic star formation history at $z\sim2$ \citep[e.g.][]{Madau1998,hopkins&beacom06}. 
Although it represents a significant epoch
in the evolution of the observable Universe, the properties of galaxies 
remain largely unexplored at this epoch, as it has only recently
become accessible by current observational facilities \citep[CANDELS, GOODS, Herschel, ALMA,][]{alma,Lin-deep, deep3, sinfoni}.
 
As a result, what constitutes arguably the most important aspect of hierarchical galaxy formation and evolution is still being debated. 
To what extent mergers, as opposed to secular evolution driven by (cold) gas inflows, explain the diversity of galaxies?
The significance
of mergers, considered a cornerstone of the bottom-up growth of galaxies, has been heavily debated
in recent work. They are certainly capable of inducing star
formation, black hole growth and morphological transformations
\citep[e.g.][]{Springel05}, but it is not obvious 
that, \emph{at $z>1$}, mergers drive the evolution of galaxy properties like stellar mass,
size and morphology \citep{Shankar04, Law09, Kaviraj13} considering the steady input of accreted streams of gas~\citep{keresetal05, ocvirketal08, dekeletal09} 
and the gas-rich nature of galaxies~\citep{tacconietal10, santinietal14}.

Observational studies 
suggest that large fractions of star-forming galaxies around
$z\sim2$ are not mergers but show kinematics and visual
morphologies that are more consistent with systems dominated by
turbulent discs~\citep[e.g.][]{Forster06,Shapiro08,Genzel08,Mancini11,Kaviraj13}.
In addition, many primordial spheroids that are forming the bulk of
their stellar mass at $z\sim2$ do not show the tidal features
that would be expected from recent major mergers~\citep{Kaviraj13b}. 
Notwithstanding these advances, large statistical studies remain difficult, both due to
the fact that observational efforts rely on pencil-beam surveys
that are susceptible to low-number statistics and cosmic variance
but also that they are based on techniques that can differ significantly from
study to study. Moreover, studies of galaxy merging at these
redshifts are further complicated by the fact that normal star
forming discs becomes more turbulent and asymmetric at earlier
times, making them difficult to separate from genuine mergers \citep{KHC14,HCK14}. 

On the other hand, semi-analytical models and numerical simulations propose that mergers can account for the size increase of \emph{local} early-type galaxies if they are mostly dry (gas poor) and minor mergers~\citep{boylankolchin06,khochfarsilk06b,maller06,naabkhochfar06,naab07,bournaudjog07,delucia07,guowhite08,hopkins09,nipoti09,feldmann10,shankar13,Bedorf13}.
Dry minor mergers explain the loss of compactness of massive ellipticals at $z<2$, where they are thought to take over smooth accretion processes in terms of stellar mass increase rates~\citep{Oser10, lackneretal12, hirschmannetal12, Dubois2013, lee&yi13}.
The dryness of low-redshift galaxies is ensured either by the environment (for satellites infalling in groups and clusters) or by the presence of a supermassive black hole (BH) at the center of massive galaxies 
which powers feedback from the active galactic nuclei (AGN)~\citep{dimatteoetal08, booth&schaye09, duboisetal12}. Together, these mechanisms allow for the formation of extended 
elliptical galaxies that would otherwise remain compact discs ~\citep{Dubois2013,choietal15}.

Multiple numerical studies also focused on a few idealised high resolution merger events to determine their impact on the morphology of the stellar component of galaxies \citep{bournaud04,bournaud05,naabt06,Peirani10}. They found that while major mergers, or multiple minor mergers of stellar disks tend to produce elliptical-like remnants, either disky or boxy depending on the amount of gas available  \citep{cretton01, naab03,naabj06,Qu11}, single minor mergers did not systematically destroy the primary disk but only thickened it  \citep{quinn93,walker96,velazquez99,younger07}. 
On the other hand, the steady input of cosmological gas accretion is able to rebuild the disc of galaxies~\citep{brooksetal09, agertetal09, pichonetal11}.
Hence one needs to asses the relative importance of mergers versus smooth accretion driven by the cosmic environment and to study its induced morphological diversity.
With the advent of large-scale albeit fairly well resolved cosmological hydrodynamical simulations such as  \hagn\!\!  (\citealp{duboisetal14}, see as well~\citealp{devriendt10, khandaietal14, vogelsbergeretal14, schayeetal15} for similar simulations performed with different numerical techniques), it has recently become feasible to investigate these different physical processes in detail and with sufficient statistics, a necessary requirement to truly unravel the impact of galaxy environment on their properties.  

\begin{figure*}
\center \includegraphics[width=2.25\columnwidth]{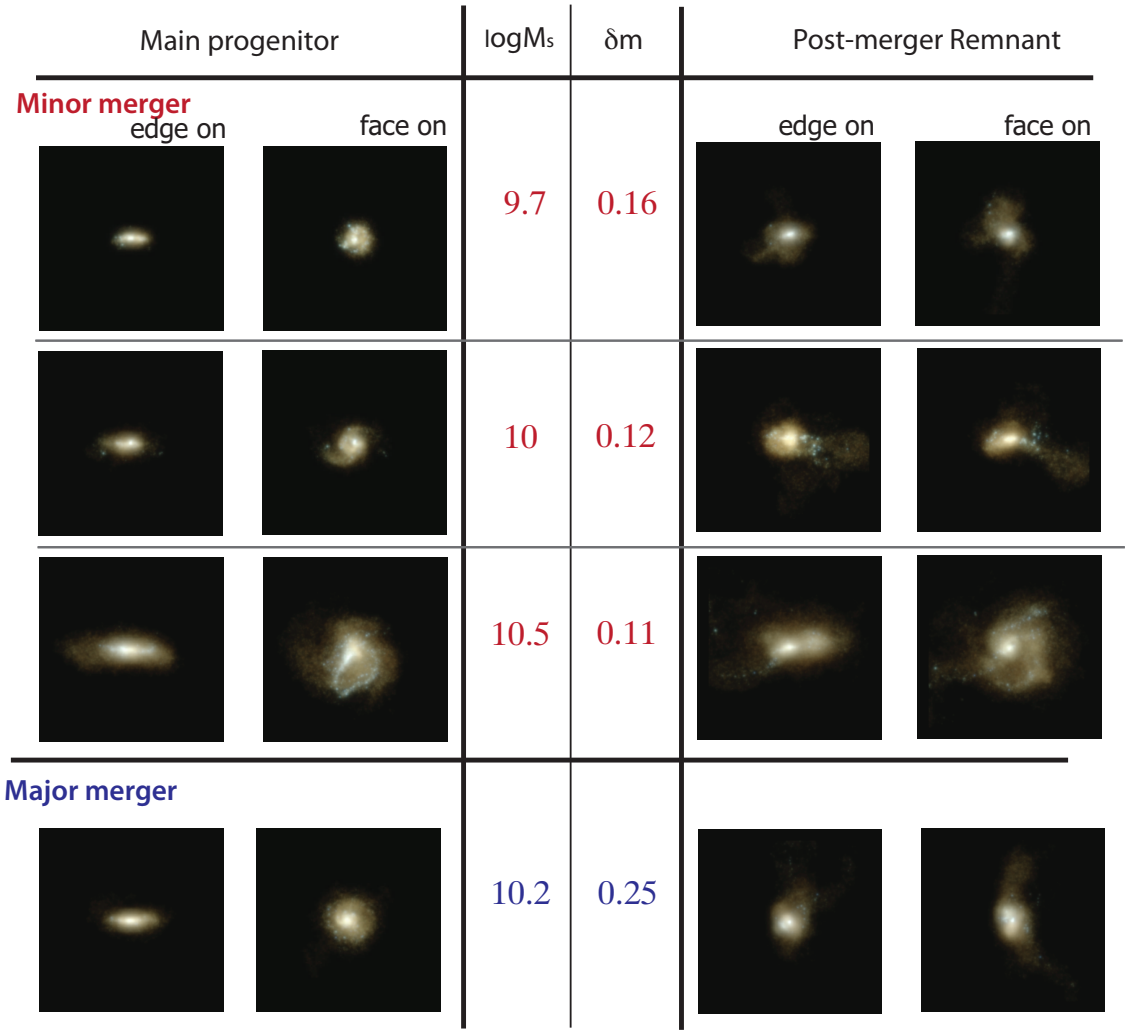}
  \caption{Rest-frame color images (\emph{u}, \emph{g} and \emph{i} filters) of a sample of Horizon-AGN disk galaxies caught during their pre-merger phase at $z=2.2$ (two leftmost columns) and their post-merger phase at $z=1.9$ (two rightmost columns). The first and third columns are edge-on views, and the second and fourth columns are face-on views. Extinction by dust is not taken into account. Each frame is 100 kpc on a side. $M_{\rm s}$ is the stellar mass of the main progenitor in $\rm M_\odot$ units and $\delta m$ is the mass ratio of the merger (see text for exact definition). This figure illustrates the ability of mergers (major but also minor) to turn disk-like galaxies into spheroids.}
\label{fig:visual}
\end{figure*}

Following up on 
 \cite{welkeretal14} which explored the fundamental role of mergers and smooth accretion on the evolution of the galactic spin and its orientation with respect to the cosmic web,
we investigate the comparative role of mergers and smooth accretion of both gas and stars on defining and modifying the size and morphology of galaxies (see Fig~\ref{fig:visual}). In the first section we describe the main features of the simulation and present the numerical analysis methods used to derive our results. In section 3, we evaluate the accretion rates of different types of mergers as well as smooth accretion over cosmic history. In section 4, we analyze the impact of both processes on the growth of galaxies in the cosmic web, with specific emphasis on the different role played by gas and stars dominated mergers (equivalent  to the dry/wet dichotomy used in lower $z$ studies).  In section 5, we explore the competitive effects of smooth accretion and mergers on the morphology of galaxies and their correlation to the disk and spheroid abundances over the duration of the peak of cosmic star formation. Our main results and conclusions are summed up in section 6.

\section{Numerical methods and definitions}
\label{section:virtual}

We briefly describe the cosmological hydrodynamical simulation used for analysis in this paper, \hagn\!\!, which is already described in more details in~\cite{duboisetal14}.
The \hagn\!\! simulation is run with a $\Lambda$CDM cosmology with total matter density $\Omega_{\rm  m}=0.272$, dark energy density $\Omega_\Lambda=0.728$, amplitude of the matter power spectrum $\sigma_8=0.81$, baryon density $\Omega_{\rm  b}=0.045$, Hubble constant $H_0=70.4 \, \rm km\,s^{-1}\,Mpc^{-1}$, and $n_s=0.967$ compatible with the WMAP-7 data~\citep{komatsuetal11}.
The size of the simulation box is $L_{\rm  box}=100 \, h^{-1}\rm\,Mpc$ on a side, and the volume contains $1024^3$ dark matter (DM) particles, corresponding to a DM mass resolution of $M_{\rm  DM, res}=8\times 10^7 \, \rm M_\odot$.
The simulation is run with the {\sc ramses} code~\citep{teyssier02}, and the initially coarse $1024^3$ grid is adaptively refined down to $\Delta x=1$ proper kpc, with refinement triggered in a quasi-Lagrangian manner: if the number of DM particles becomes greater than 8, or the total baryonic mass reaches 8 times the initial DM mass resolution in a cell.
It lead to a typical number of $6.5\times 10^9$ gas resolution elements (leaf cells) in the \hagn\!\! simulation at $z=1$.
Heating of the gas from a uniform UV background takes place after redshift $z_{\rm  reion} = 10$ following~\cite{haardt&madau96}. 
Gas can cool down to $10^4\, \rm K$ through H and He collisions with a contribution from metals using rates tabulated by~\cite{sutherland&dopita93}. 
Star formation occurs in regions of gas number density above $n_0=0.1\, \rm H\, cm^{-3}$ following a Schmidt law: $\dot \rho_*= \epsilon_* {\rho_{\rm g} / t_{\rm  ff}}$,  where $\dot \rho_*$ is the star formation rate mass density, $\rho_{\rm g}$ the gas mass density, $\epsilon_*=0.02$ the constant star formation efficiency, and $t_{\rm  ff}$ the local free-fall time of the gas.
Feedback from stellar winds, supernovae type Ia and type II are included into the simulation with mass, energy and metal release.
The simulation also follow the formation of black holes (BHs), which can grow by gas accretion at a Bondi-capped-at-Eddington rate and coalesce when they form a tight enough binary.
BHs release energy in a quasar/radio (heating/jet) mode when the accretion rate is respectively above and below one per cent of Eddington, with efficiencies tuned to match the BH-galaxy scaling relations at $z=0$~\citep[see][for details]{duboisetal12agnmodel}.

Galaxies are identified with the most massive sub-node method~\citep{tweedetal09} of the AdaptaHOP halo finder~\citep{aubertetal04} operating on the distribution of star particles with the same parameters than in~\cite{duboisetal14}. 
Unless specified otherwise, only structures with a minimum of $N_{\rm  min}=100$ particles are considered, which typically selects objects with masses larger than  $2\times10^8 \, \rm M_\odot$. Catalogues containing up to $\sim 150 \, 000$ galaxies are produced for each redshift output analysed in this paper ($1.2 < z < 5.2$). Note that, although sub-structures may remain, these populations of galaxies are largely dominated by main structures.

The galaxy  catalogues are then used as  an input to build merger trees with TreeMaker~\citep{tweedetal09}. 
Any galaxy at redshift $z_n$ is connected to its progenitors at redshift $z_{n-1}$ and its child at redshift $z_{n+1}$. 
We build merger trees for 22 outputs from $z=1.2$ to $z=5.2$ equally spaced in redshift.
On average, the redshift difference between outputs corresponds to a time difference of 200 Myr (range between 100 and 300 Myr).
We reconstruct the merger history of each galaxy (halo) starting from the lowest redshift $z$ and identifying the most massive progenitor at each time step as {\it the galaxy} or {\it main progenitor}, and the other progenitors as {\it satellites}. Moreover, we  check that the mass of any child contains at least half the mass of its main progenitor to prevent misidentifications.
Note that such definition of mergers (vs smooth accretion) depends on the threshold used to identify objects as any object below the chosen threshold is discarded and considered as smooth accretion. 

We sort mergers in three categories depending on the mass fraction $\delta m = m_{\rm mergers} (z_{n-1 \rightarrow n })/ M_{s}(z_{n})$, where $m_{\rm mergers}(z_{n-1 \rightarrow n})$ is the stellar mass accreted through mergers between $z_{n-1}$ and $z_{n}$ and $M_{s}(z_{n})$ the stellar mass of the merger product at $z_{n}$. Major mergers are defined as mergers with $\delta m>20\%$, minor mergers as mergers with $9\%<\delta m<20\%$ and very minor mergers with  $4.5\%<\delta m<9\%$.  Any merger with $\delta m<4.5\%$ is discarded and counted as smooth accretion. These bins are defined so as to be consistent with observational definitions of mergers using pairs of interacting galaxies, for which the observed mass ratio $R$ is defined as $R=M_{\rm satellite}/M_{\rm galaxy}$ and where the subscripts indicate secondary and main progenitors respectively, as defined in the previous paragraph of this section. Our bins thus correspond to $R=1:4$, $R=1:10$ and $R=1:20$.

\begin{figure}
\center \includegraphics[width=1\columnwidth]{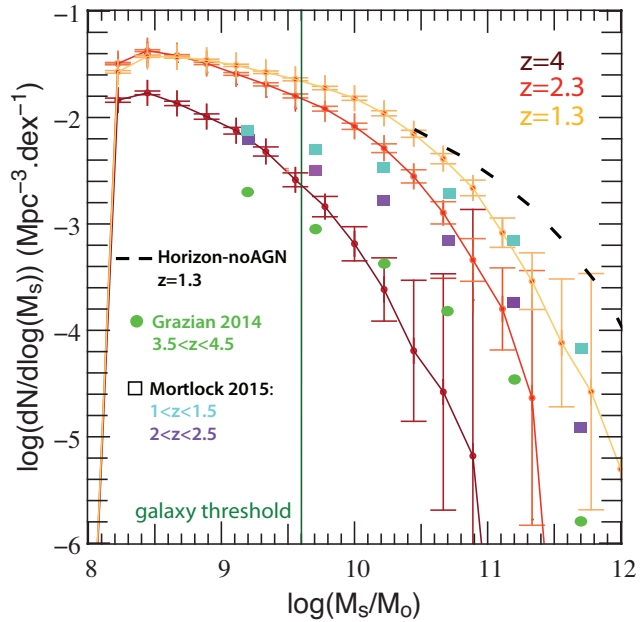}
 \caption{Galaxy stellar mass function in Horizon-AGN, for $z=4$ to $z=1.3$. $N$ is the number density of galaxies, $M_{s} $ the stellar mass (together with Horizon-noAGN for comparison). The sharp cut-off at $M_{s}=10^{8} \, {\rm M_\odot}$ corresponds to our completeness detection threshold. Observational points from CANDELS-UDS and GOOD-S surveys are rescaled from best fits in \citet{Mortlock15} and \citet{Grazian14} and overplotted. While mass functions are consistent at the high mass end, Horizon-AGN overshoots the low-mass end by about a factor 3 in this redshift range. The vertical green line shows the selection threshold  for our main progenitors candidates, chosen to enable us to completely track their mergers with galaxies up to 20 times smaller. }
\label{fig:mass_function}
\end{figure}

In order to preserve completeness, we define a second threshold and exclude galaxies with $M_{\rm  s} <5\times10^9 \, \rm M_\odot$ from the {\it galaxy} sample used in our analysis. {\it Satellites}, however, are allowed to be less massive, in order for us to still capture the whole range of merger ratios for the smallest of our galaxies. The {\it galaxy} threshold is identified as the green vertical line on Fig.~\ref{fig:mass_function}, which displays the mass function for all the structures identified in the simulation (for comparison we have also plotted the mass function of the Horizon-noAGN simulation, which is the same simulation performed without BHs and, therefore, AGN feedback). As can be seen on the figure our sample is complete down to our strict selection threshold of $M_{s}=2\times 10^{8}\, {\rm M_\odot}$, corresponding to galaxies with 100 star particles. Thus, the smallest mergers detectable for galaxies at the $M_{\rm  s} <5\times10^9 \, \rm M_\odot$ threshold correspond to a mass ratio of $\delta m=4.5\%$, which means that our merger classification is complete for our {\it galaxy} sample.

The gas content and its properties (density, metallicity, pressure, temperature) of each galaxy is extracted from the AMR grid, considering all cells within its effective radius. (see Appendix~\ref{sec:gas} for the technical subtleties).

Since the gas needs to be cold and dense enough to be eligible to form stars, we define ``cold'' gas (in the sense of star forming gas) the cells with a gas density higher than $n>0.1\, {\rm H\, cm}^{-3}$ and a temperature $T\leq10^4\, \rm K$. We also define the gas fraction $f_{\rm gas}$ of a galaxy as $f_{\rm gas} = M^{\rm cold}_{\rm gas}/(M_{0.5} + M^{\rm cold}_{\rm gas})$ with $M^{\rm cold}_{\rm gas}$ the mass of cold gas and $M_{0.5}$ the mass of stars, both enclosed within the sphere of radius $r_{0.5}$. As can be seen in Fig.~\ref{fig:gas_fraction}, this quantity decreases with stellar mass and redshift due to star formation and feedback, older galaxies becoming more massive after they used the gas available to form stars and/or after it has been blown out of them by AGN/supernova feedback. This evolution is consistent with previous numerical studies \citep[e.g.][]{duboisetal12agnmodel, popping14}.

\section{Mergers and smooth accretion across the peak of cosmic star formation history}
\label{section:rates}

While it is now well established that mergers have a significant impact on $z<1$ early-type galaxy sizes and kinematics (see references in introduction), it is not yet clear whether (i) this extends to the galaxy population at high redshift and (ii) over which timescale they are of importance, as many galaxies may not merge at all for long periods of time. Observations of local galaxies not only suggest that early-type galaxies increased their size by 3-5 from $z \sim 2$, but also that, while most massive ones ($M_{s}> 1.5 \times 10^{11} \, \rm M_\odot$) roughly doubled their size from $z \sim 1$, smaller ones underwent a more limited growth between $z \sim 1$ and z=0 (by a factor 1.1 to 1.3, \citealp{Huertas13b}).
From these results, we expect a growth by at least a factor $2-2.5$ between $z \sim 2-3$ and $z=1$ \citep{Nipoti12}.
To quantify the relative contribution of mergers and smooth accretion to the total mass budget of galaxies over the range of redshifts corresponding to the peak of cosmic star formation history down to $z=1$, we therefore compute the rates of galaxies having undergone at least a merger within our mass fraction bins at these redshifts. 

\begin{figure*}
 \includegraphics[width=1.8\columnwidth]{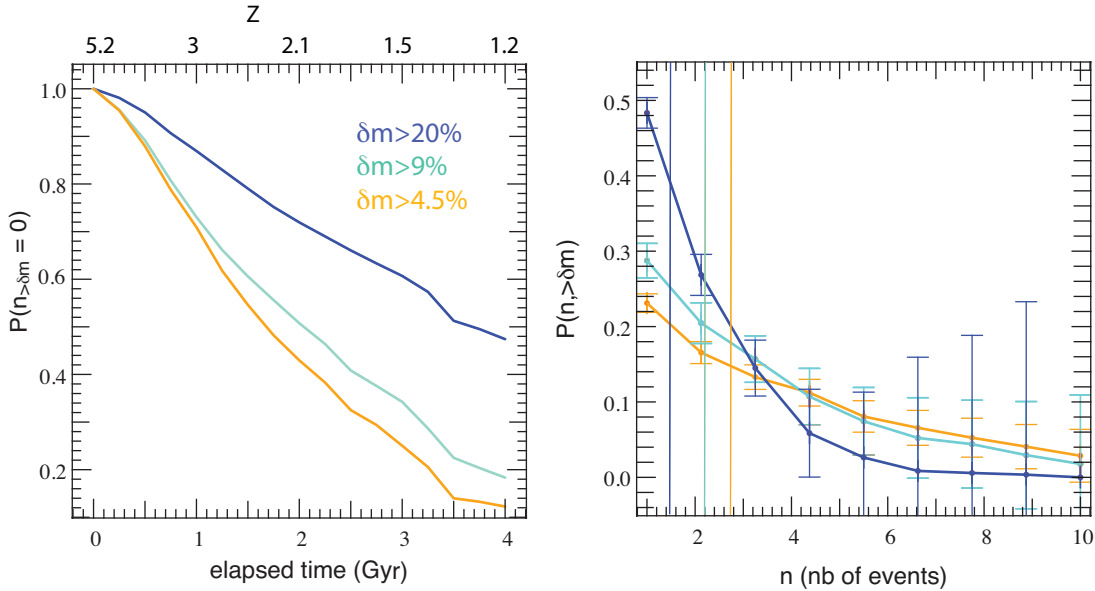}
  \caption{\emph{Left panel}: evolution of a sub-sample of galaxies identified at $z=1.2$ with $M_{\rm s}>10^{10}\, {\rm M_\odot}$ and which can be tracked to $z= 5.2$. This panel shows the probability for galaxies 
\emph{not} to undergo a major or minor merger during the redshift interval. \emph{Right panel}: PDF of the number $n_{m}$ of galaxy mergers of a given mass ratio, $\delta m$, undergone between redshifts $5.2 \geq z \geq 1.2$. This PDF is restricted to galaxies with at least one very minor merger. Vertical lines show the average value for each sample. It illustrates the paucity of major mergers: most galaxies which merged, have had at most one major merger across this cosmic time interval, while they go through on average 2 to 3 mergers. Note the large variations from one galaxy to another.}
\label{fig:rates}
\end{figure*}

We find that, at $z=1.2$, around $35\%$ of galaxies with $M_{\rm s}>10^{10}\, {\rm M_\odot}$ have undergone at least one major merger, $80\% $  a minor merger, and $85\% $ a very minor merger. These results are consistent with  findings by \cite{Kaviraj14}, (our minor merger rates are slightly inferior due to a coarser redshift  sampling).
Fig.~\ref{fig:rates} (left panel) presents the evolution of those rates with redshift, focusing on the sub-sample of galaxies in this mass range who possess a progenitor at $z=5.2$ (sub-sample of 15 000 galaxies). It displays the evolution of the fraction of this sub-sample which remains free from mergers of a given type (major, minor and very minor) as a function of cosmic time. It shows that over this 4 Gyr period, $\sim 50\% $ of the sample undergoes a major merger and therefore that mergers, especially minor ones, are quite frequent
over the whole redshift range. The sample is affected by mergers at an average rate around $1-2\times10^{-3} \, {\rm Gyr}^{-1} \, h^{3} \, {\rm Mpc}^{-3}$ and $3-5\times 10^{-4} \, {\rm Gyr}^{-1}\, h^{3}\, {\rm Mpc}^{-3}$  for all mergers and major mergers respectively. Note that these values are completely consistent with observations by \cite{Lotz11} and in good agreement with the cumulative merger rates per galaxy derived from the Illustris simulation \citep{Illustris-rates} .
 
The right panel of Fig.~\ref{fig:rates} focuses on galaxies which have had at least a merger between $z=5.2$ and $z=1.2$. It shows the probability distribution function (PDF) $P(n, >\delta m)$ for these galaxies  to have undergone a number $n$ of mergers of a given mass ratio, $\delta m$,  between $z=5.2$ and $z=1.2$. This PDF indicates that, while most of these galaxies underwent at most a single major merger, on average they undergo two to three merger events.

\begin{figure*}
\includegraphics[width=0.86\columnwidth]{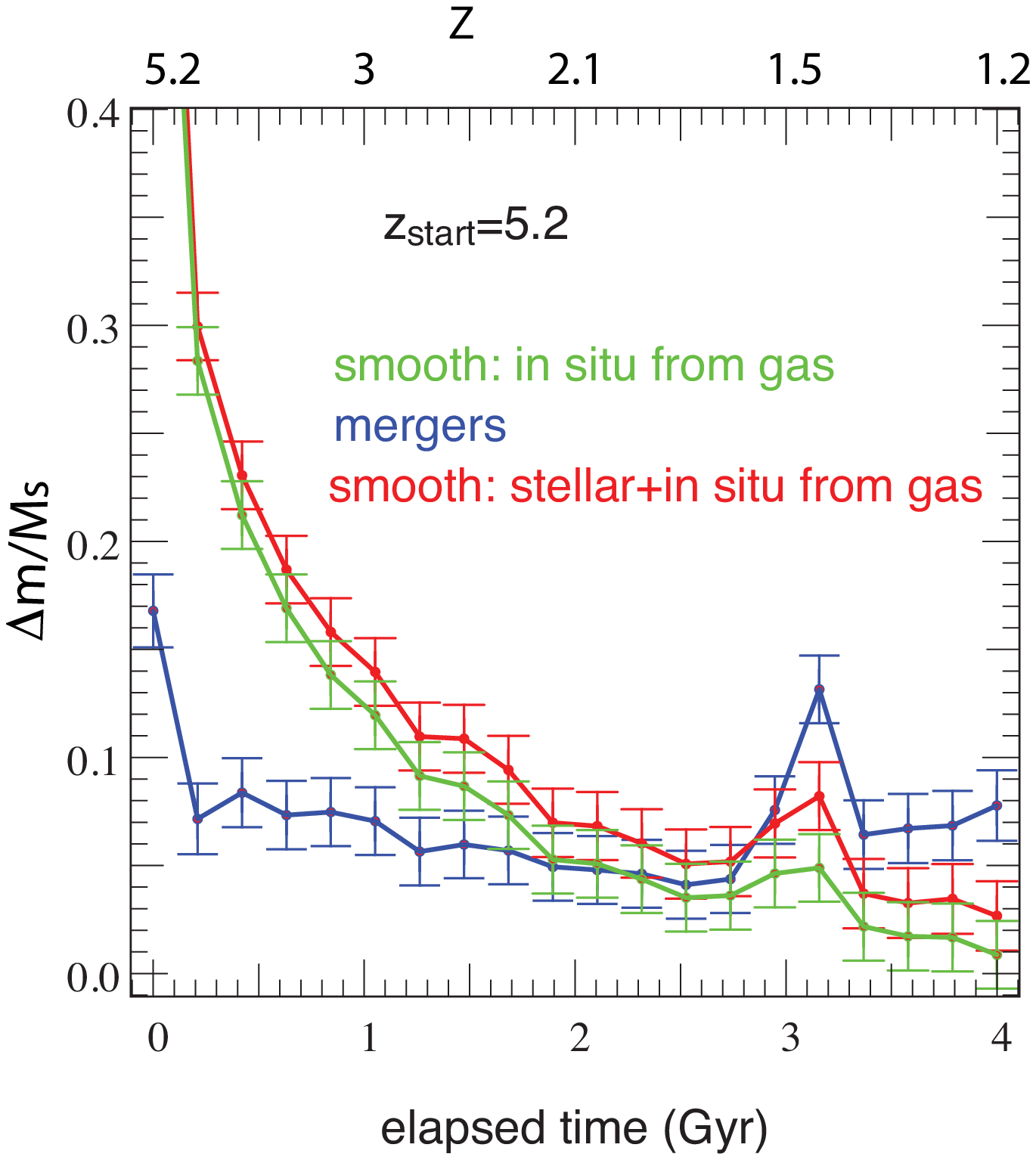}\hskip 0.1cm
\includegraphics[width=0.9\columnwidth]{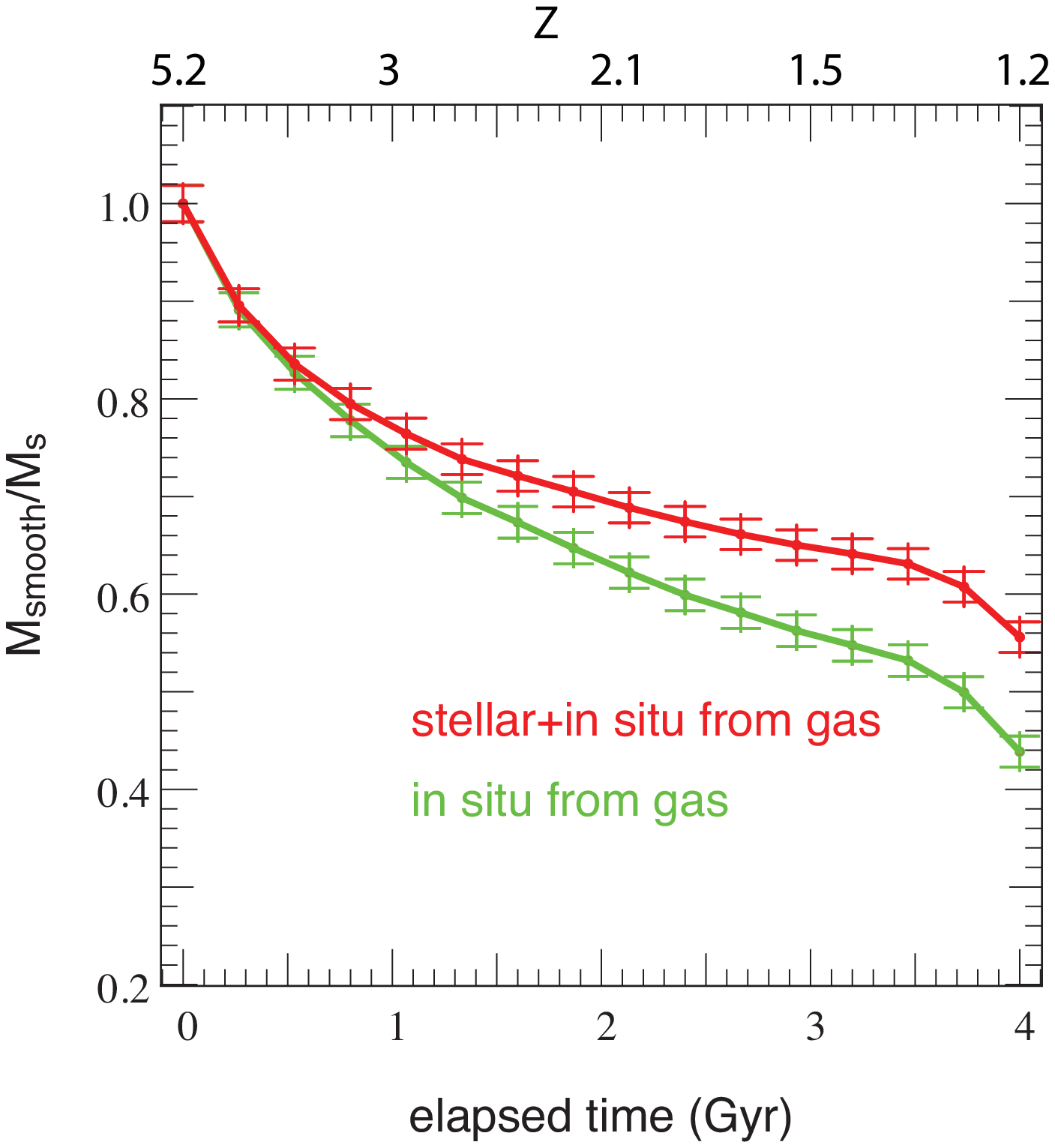}
  \caption{\emph{Left panel}: evolution of $\Delta m/M_{s}$ over cosmic time for galaxies with $M_{s} > 10^{10} \, {\rm M_\odot}$ at $z=1.2$, where $\Delta m$ is the mass increase due to in situ formed stars (green), merger with a companion (blue), or  in situ formed stars combined with the 'diffuse' accretion of stars (i.e. stars not identified as belonging to any galaxy, red) between two time steps ($\sim 200$ Myr), and $M_{s}$ is the stellar mass. We plot average values for all selected galaxies at every time output. Note that mergers and smooth accretion contribute similarly to the mass growth of galaxies from $z=3$ onwards.
\emph{Right panel}: evolution of $M_{\rm smooth}/M_{s}$ over cosmic time from $z=5.2$ to $z=1.2$, where $M_{\rm smooth}$ is the mass of stars either produced in situ from the gas component, or accreted 
 'smoothly' (i.e. star particles not associated with a galaxy above our mass resolution threshold), for galaxies with $M_{s} > 10^{10} \, {\rm M_\odot}$ at $z=1.2$. At $z=1.2$, about half of the stellar mass of these galaxies comes from such smooth accretion processes.}
\label{fig:rates_smooth}
\end{figure*}

Fig.~\ref{fig:rates_smooth} reveals that both mergers and smooth accretion ought to be taken into account to attempt to understand the morphology distribution of galaxies. This figure shows the evolution of the smooth accretion and merger contributions to the mass budget of galaxies across the peak of cosmic star formation history. The first panel presents the evolution of  $\Delta m/M_{s}$ averaged for all galaxies with $M_{s} > 10^{10} \, {\rm M_\odot}$ at $z=1.2$, with $\Delta m$ the mass accreted between two successive time steps (i.e. over a period of $\sim200$ Myr) and $M_{s}$ the stellar mass. The red and green curves correspond to the mass fraction accreted via smooth accretion of gas (green curve) and stars (red curve: gas+stars; i.e. including mergers with $M_{s} < 2 \times 10^8 \, {\rm M_\odot}$ galaxies), and the blue curve corresponds to the mass fraction accreted through mergers. While at high-redshift ($z \sim 5$) young and small galaxies undergo a rapid relative mass growth through accretion of gas and swift merging of very small structures close to the detection threshold, this activity settles around $z \sim 3-4$, when effects of smooth accretion and mergers on mass growth become comparable, until mergers slightly take over around $z \sim 1.5$. The net result
is that at $z=1.2$, $\sim 45\%$ of the galactic stellar mass can be attributed to in situ formation from smooth accretion of gas, as can be seen on Fig.~\ref{fig:rates_smooth} (second panel) .

In conclusion, mergers and smooth accretion contribute equivalently to the galactic mass budget over the peak of cosmic star formation history. 
It therefore seems that in order to understand the evolution of galactic sizes and morphologies over this period, one needs to account for the 
possibility that these two processes play different roles. This is what we explore in the next sections.

\section{Growth of galaxy sizes }
\label{section:growth} 

\begin{figure*}
\includegraphics[width=2.2\columnwidth]{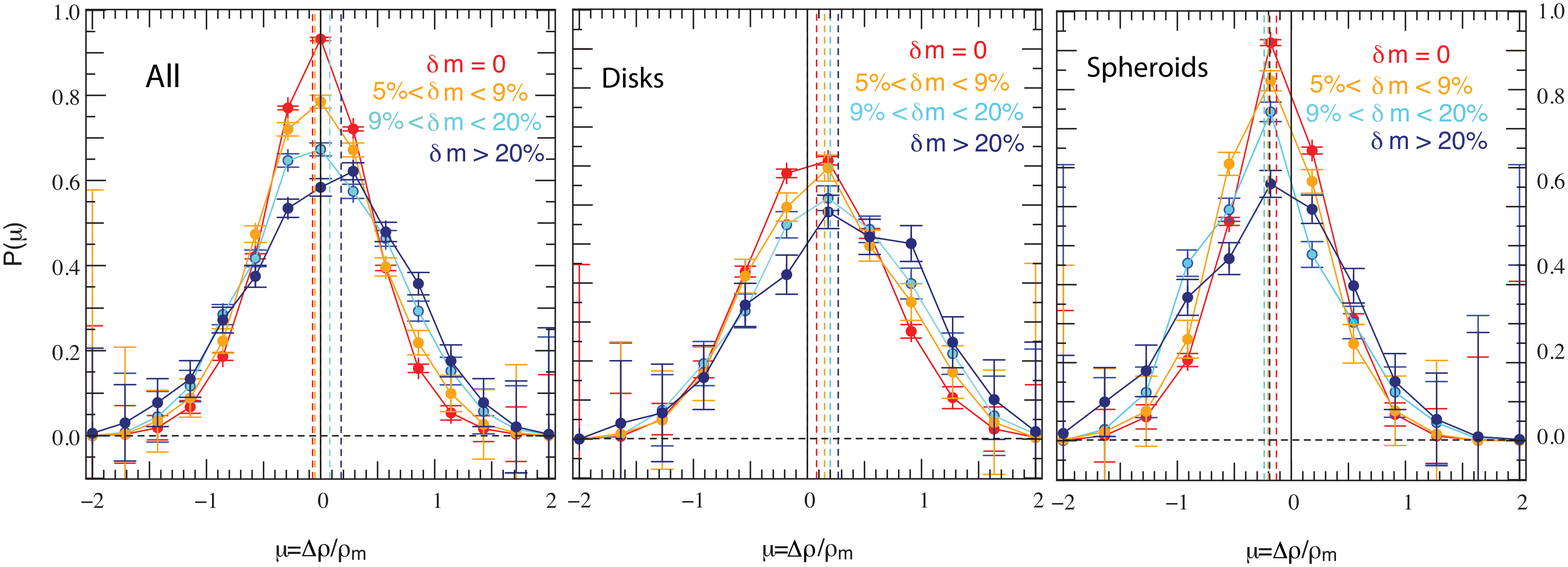}
  \caption{\emph{Left panel:} PDF of the density growth ratio $\mu=2(\rho_{n+1}-\rho_{n})/(\rho_{n}+\rho_{n+1})$ , where $\rho_{n}$ is the density of the galaxy within its half mass radius at time step $n$, for different merger mass ratios. This ratio is calculated for galaxies with $M_s>10^{9.5} M_{\odot}$ over each time step between $1.2 \leq z \leq 5.2$, and all these timesteps are then stacked. Each vertical dashed line shows the average value for the merger mass ratio bin of the corresponding color. Mergers have a tendency to widen the distribution and increase the stellar density, especially major mergers (vertical dashed line on the positive side of $\mu$ values). However, this behavior is actually different for galaxies which are disks prior to the merger (for which the stellar density rises: \emph{middle panel}) and for those which are originally spheroids (for which the stellar density decreases: \emph{right panel}).}
\label{fig:vol_mass}
\end{figure*}
%
Fig.~\ref{fig:vol_mass} shows the evolution of the stellar density, obtained by adding the masses of all star particles enclosed within the half mass radius of the galaxy. Since the shape of galaxies can vary significantly over the cosmic time interval spanned by our study, especially when galaxies merge, we take anisotropy into account. More specifically, the density $\rho$ is  defined as $\rho=3 M_{0.5}/(4 \pi a b c)$ with $a>b>c$ the lengths of the semi-principal axes of the galaxy derived from the eigenvalues of the inertia tensor, as explained in Appendix~\ref{sec:math}, and $M_{0.5}$  is the sum of all the masses of the star particles contained within its half mass radius. The left panel of the figure shows the PDF of  the relative density growth $\mu=2(\rho_{n+1}-\rho_{n})/(\rho_{n}+\rho_{n+1})$, where $\rho_{n}$ is the average density of the galaxy within its half mass radius at time step $n$, stacked for each time output of the simulation between $1.2 \leq z \leq 5.2$. Notice how mergers tend to widen the distribution, populating the high-compactions and high-dilatation tails of the distribution.  Looking at this panel, one might think that smooth accretion and very minor mergers tend to lower the density of the merger remnant on average, while minor and major mergers tend to increase it, but it is actually highly dependent on the initial morphology of the galaxy. This can be seen on the middle and right panels of Fig.~\ref{fig:vol_mass}. Galaxies that are initially disks show an increased stellar density after mergers, the effect being stronger the higher the mass ratio of the merger. On the other hand, galaxies which begin as spheroids tend to have their stellar density decreased by mergers, the effect being statistically stronger for minor mergers. $55\%$ of spheroids which merge betray a decrease in stellar density after $\sim$ 200 Myr, and only 40$\%$ of the non-merger galaxies do the same. It is interesting to notice that, in both cases, this increase/decrease in average density is related to how skewed the distribution becomes and not only to a global drift towards positive/negative values. As a result, minor and major mergers of spheroids are much more likely to trigger important decreases in stellar density (by more than a factor 2) than smooth accretion: $16\%$ of cases versus $5\%$ respectively, and even as low as $3\%$ if events where stars are accreted below our galaxy mass threshold are discarded. Around $8\%$ of major mergers and even fewer minor mergers trigger dilatations by more than a factor 5.

Similarly, $73\%$ of disk galaxies increase their density after merging (against $63\%$ for non-mergers). However, and more importantly, $30\%$ of these mergers increase it by at least a factor 2 as compared to only $9\%$ for smooth accretion. Finally, only $10\%$ of the major mergers and the minor mergers lead to compactions by more than a factor 5.

These results statistically support the claim that mergers turn disks into denser structures while they tend to lower the density of spheroids. Moreover, although major mergers are found to be quite rare, the ability of the more frequent minor mergers to trigger effects of comparable amplitude points towards an important role of multi-minor mergers in driving the size-mass relationship of early-type galaxies~\citep{Kaviraj14b}.

\begin{figure*}
\center{\includegraphics[width=1.9\columnwidth]{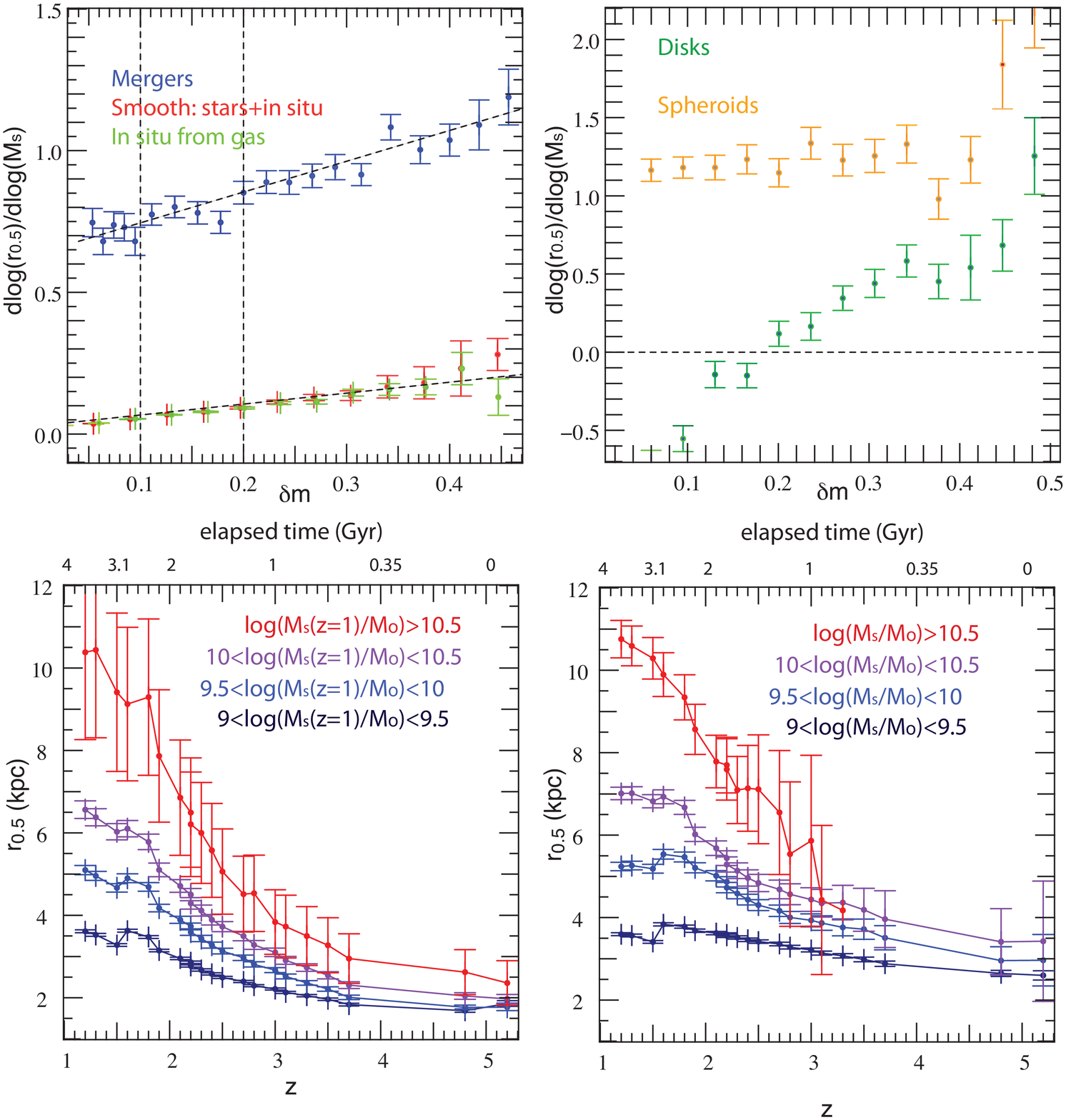}}
  \caption{\emph{Top left panel}: evolution of the logarithmic derivative of the half-mass radius $r_{0.5}$ with respect to the mass ratio $\delta m=\Delta m/M_{\rm s}$  ($\Delta m$ is the stellar mass gained between two consecutive outputs) of the merger or smoothly accreted material for all galaxies and all time outputs between $z=5.2$ and $z=1.2$. Filled blue symbols indicate mergers, red and light green ones represent smoothly accreted mass, including a stellar component (red) or gas only (green). Whilst the evolution is linear in each case, the dependence of radius growth on the mass ratio is found to be much steeper for mergers. \emph{Top right panel}: same plot as the top left panel, except we have split the merger sample according to different pre-merger morphologies: disks (green) and spheroids (yellow). The steepness of the radius versus $\delta m$ relation appears mainly caused by minor merger disruption of the disks. \emph{Bottom left panel}: evolution of the half-mass radius as a function of redshift for galaxies split into bins of different mass at redshift $z=1.2$. \emph{Bottom right panel}: same plot as in the bottom left panel but for galaxies split into redshift independent mass bins.}
\label{fig:dr_minor_major}
\end{figure*}

We further analyze the role of mergers in driving galaxy stellar density evolution by looking at the relation between growth in stellar mass and growth in stellar half-mass radius.
To do so, we compute the evolution of the logarithmic derivative of the half mass radius $r_{0.5}$ with respect to $M_{\rm s}$ as a function of the mass ratio $\delta m=\Delta m/M_{\rm s} \propto  \Delta \log_{10} M_{\rm s}$, where $\Delta m$ is the stellar mass accreted between two consecutive outputs through mergers (blue curve) or smooth accretion (red and green curves) for all galaxies and all time outputs between $z=5.2$ and $z=1.2$. Note that $M_{\rm s}$ or $M_{0.5}$ are equivalent for the purpose of this comparison so we use them interchangeably. The result is shown in the top left panel of Fig.~\ref{fig:dr_minor_major}.  While mergers and smooth accretion drive a similar amount of mass growth (see the left panel of Fig.~\ref{fig:rates_smooth} in the previous section), mergers are much more efficient drivers of galaxy size growth. Ignoring the very shallow dependence on $\delta m$, smooth accretion processes lead to an average radius-mass relation $r_{0.5} \propto M_{s}^\alpha$ with $\alpha = {0.1 \pm 0.05}$. Slightly higher values of $\alpha$ are reached  for higher $\delta m$ (see left panel of Fig.~\ref{fig:dr_minor_major}), but always remain within a factor 2, i.e. $\alpha \leq 0.2$. This dependence on $\delta m$ can be explained by the fact that higher mass ratio values most often correspond to lower stellar masses, for which cold flows bring in more specific angular momentum than hot phase accretion in more massive galaxies \citep{kimmetal11}.
For mergers, we obtain $r_{0.5} \propto M_{s}^\beta$ with $\beta = 0.85 \pm 0.3$. $\beta$ also increases with $\delta m$ up to values $\sim 1.2$. These values are consistent with observations~\citep[e.g.][who found a value of $\beta$ around $\simeq0.6-0.8$ for early-type galaxies]{newmanetal12, cimattietal12, Huertas13b, vanderwaltetal14}, and together with the smaller values of $\alpha$, support the idea that the size growth of galaxies is mostly driven by mergers~\citep{boylankolchin06,nipoti09,feldmann10,Dubois2013}.
The bottom panels show the evolution of the average half-mass radius $r_{0.5}$ as a function of redshift when we split our galaxy sample in bins of "final" mass (i.e. galaxy masses at redshift $z=1.2$; bottom left panel) and in bins of constant stellar mass (i.e. independent of redshift; bottom right panel). Once again, the results
shown on Fig.~\ref{fig:dr_minor_major} are consistent with the overall evolution of the size-mass relationship from observations such as \cite{Huertas13b} (though our simulated galaxies are a factor of $\sim2$ larger): galaxies of a given stellar mass display much larger radii at $ z=1.2$ than their counterparts of similar stellar mass at $z=5.2$, most of the growth taking place between $z=3$ and $z=1.2$. More specifically, at $z=1.2$ galaxies with $M_{s}>10^{10.5} \, \rm M_{ \odot} $ display an average half-mass radius twice to 3 times bigger than their counterparts at $z=3$. (see bottom right panel), while one can see on the bottom left panel that galaxies reaching $M_{s}>10^{10.5} \, \rm M_{ \odot} $ at $z=1.2$ have also seen their half-mass radius grow by a factor 2 to 3 since $z=3$, and by a factor 4 since $z=5.2$.

\begin{figure*}
\includegraphics[width=1.8\columnwidth]{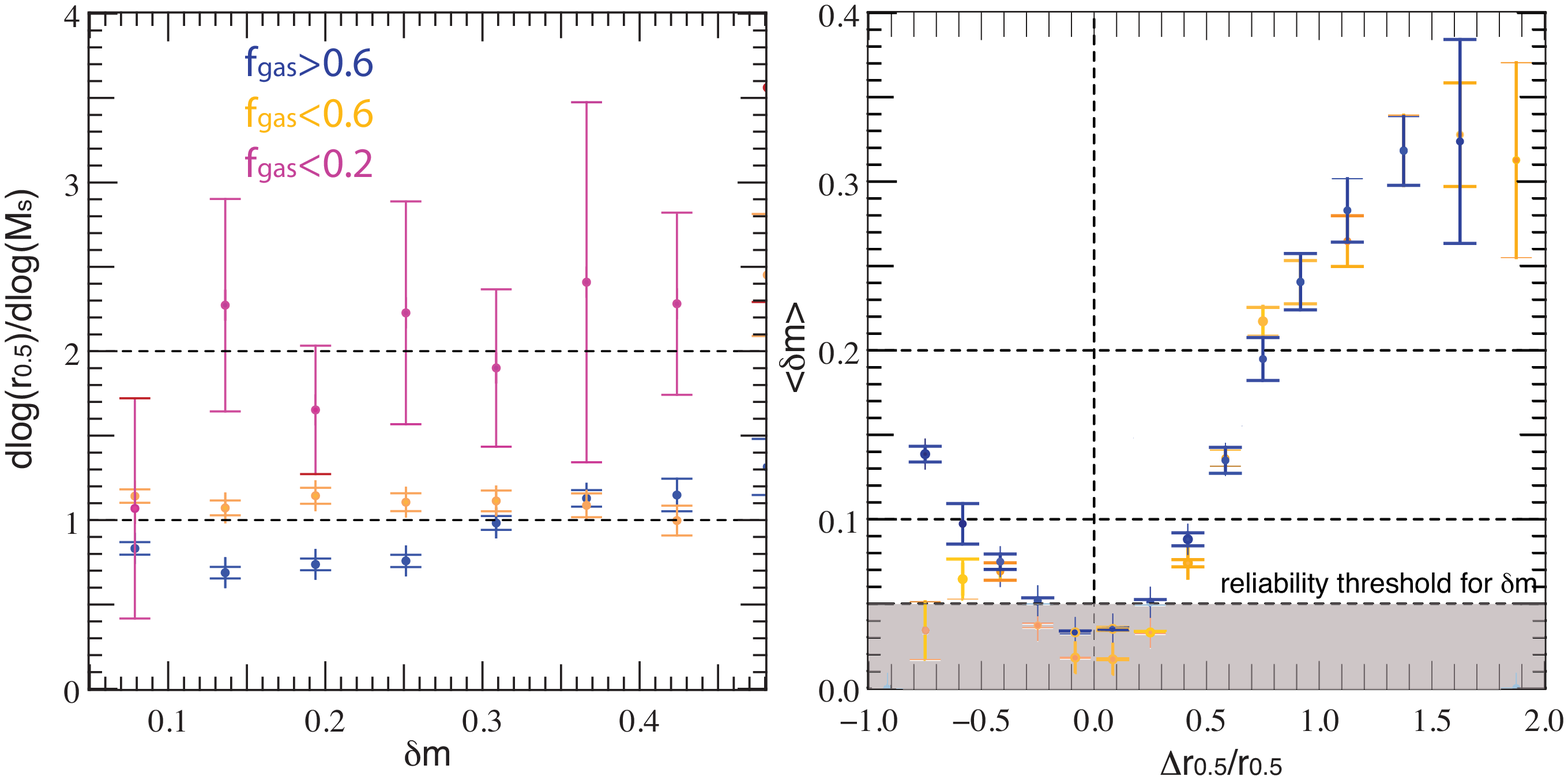}
  \caption{\emph{Left panel}: evolution of the relative variation of the half-mass radius as a function of the merger mass ratio for all galaxies which undergo a merger between $z=5.2$ and $z=1.2$ and for different pre-merger gas fractions: $f_{\rm gas}<0.2$ (pink symbols), $f_{\rm gas}<0.6$ (yellow symbols) and $f_{\rm gas}>0.6$ (blue symbols). The error bars plotted correspond to $1\sigma$ errors. 
Horizontal dashed lines represent $r_{0.5} \propto M_{\rm s}^\gamma$, with $\gamma = 1$ and $\gamma = 2$ which are predicted size-mass relations for (dry) major and minor mergers using the virial theorem \citep{Hilz2012,Dubois2013} .   Note how the presence of gas limits the radius growth. \emph{Right panel}: average mass ratio versus relative variation of the half-mass radius $\Delta r_{0.5}/r_{0.5}$ for mergers with gas fraction $f_{\rm gas}<0.6$ (yellow) and $f_{\rm gas}>0.6$ (blue). Horizontal dashed lines show major/minor/smooth accretion separation thresholds in $\delta m$. The vertical dashed line indicates the border between expansion (positive values) and contraction (negative values). Note how radius contraction is confined to wet minor mergers.}
\label{fig:dry_wet_radius}
\end{figure*}

However, Fig.~\ref{fig:dr_minor_major} does not distinguish gas-rich and gas-poor mergers. This could be potentially important as gas-poor mergers are known to trigger intense size growth of local 
early type galaxies~\citep[e.g.][]{naab07,feldmann10}, whereas accretion of gas (by gas-rich mergers or smooth accretion) is thought not to be able to since their gas shocks radiatively  and loses angular momentum, therefore piling up in the central region of the galaxy where it rapidly turns into stars and causes size contraction.
 The right panel of Fig.~\ref{fig:dry_wet_radius}, lends statistical support to this claim. It shows the average value of the merger mass ratio $\delta m$ which leads to a given relative variation of the half mass radius $\Delta r_{0.5}/r_{0.5}$, for mergers with $f_{\rm gas}>0.6$ (blue data points) and $f_{\rm gas} <0.6$ (yellow data points). From this data, one can see that radius contraction (negative values of $\Delta r_{0.5}/r_{0.5}$) is confined to gas rich minor mergers (blue data points with $ 0.09 < \delta m < 0.2$). The corresponding yellow data points for gas poor mergers are below -- or very close to -- the smooth accretion threshold (lower horizontal dashed line), indicating that smooth accretion of gas is in fact the leading advection process in those cases. Interestingly enough, major mergers $\delta m > 0.2$ 
statistically never lead to a compactification of galaxies, regardless of whether they are gas rich or not: the violent disruption that they occasion does not translate into a funelling 
of material to the central region as it does for minor mergers, but as an extended redistribution of it. 
Note that the threshold of $f_{\rm gas} = 0.6$ is chosen high compared to the values traditionally used to define wet and dry mergers at low redshift
because galaxies are more gas rich on average in the redshift range of this study. One can get an idea of how much smaller the sample gets when this 
threshold is lowered to $f_{\rm gas} = 0.2$ by looking at the left panel of Fig.~\ref{fig:dry_wet_radius}.

This panel presents the dependence of the logarithmic derivative of the half mass radius $r_{0.5}$  on the mass ratio $\delta m$ for our sample of galaxies split into different pre-merger gas fraction bins. One can see that star rich mergers with $f_{\rm gas}<0.6$ (in yellow), especially minor ones ($0.09 <\delta m < 0.2$) induce a more efficient radius growth than their gas dominated counterparts of similar mass ratio (in blue). Gas deprived mergers with  $f_{\rm gas}<0.2$ (pink curve), whether major or minor, lead to a rapid growth of the effective radius compatible with $r_{0.5} \propto M_{s}^\gamma$ where $\gamma = {2 \pm 0.5}$. This power law index is in excellent agreement with predictions from \cite{Hilz2012}, and consistent with previous numerical studies \citep{boylankolchin06, nipoti09,feldmann10}, although slightly higher, which therefore lends extra support to the scenario involving multiple dry mergers to explain the loss of compacity of massive early-type galaxies at low redshifts.

Going back to the top right panel of Fig.~\ref{fig:dr_minor_major}, we see that the dependence of the size-mass relationship on merger mass ratio can be interpreted as a morphological effect:  galaxies that are spheroids prior to the merger (yellow data points) systematically grow in size almost indistinctively with mass ratio (except for the most extreme major mergers), whilst disks (green data points) exhibit a size growth proportional to the accreted mass ratio over the same range in $\delta m$. Note that~\cite{vanderwaltetal14} find a different size-mass evolution for early-type and late-type galaxies with $\beta\simeq0.75$ and $\beta\simeq0.22$ respectively with negligible evolution with redshift. 
In our simulation, we find that spheroids (i.e. early-type galaxies) have $\beta=1.2$ on average and disks (i.e. late-type galaxies) have $\beta\simeq-0.5$ for low values of $\delta m$ and $\beta\simeq0.5$ for large values of $\delta m$, which shows a similar discrepancy of the size-mass evolution between different galaxy morphologies to that observed in~\cite{vanderwaltetal14}.
This stresses the need to study the morphology of our galaxies in further detail, and we now turn to this issue.

\section{Impact on galaxy morphologies}
\label{section:morpho} 

\begin{figure*}
\center \includegraphics[width=1.95\columnwidth]{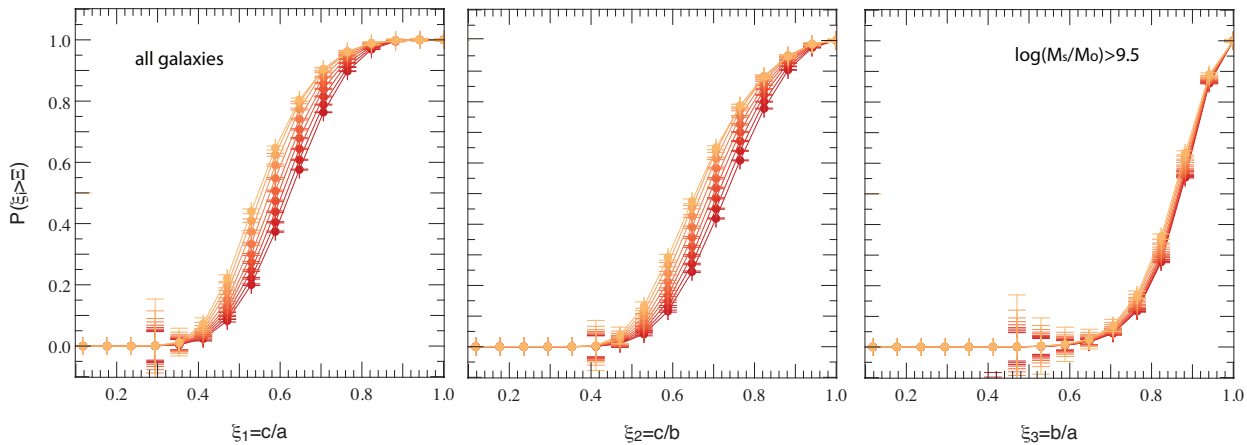}
  \caption{Cumulative PDFs of the principal semi-axis ratios $\xi_i$ for all galaxies with $M_{s}>10^{9.5}M_{\odot}$ which do not merge. Colors represent  evolution with cosmic time from $z=5.2$ (dark red) to $z=1.2$ (light orange) with an average time step of $\sim$ 200 Myr. The $1 \sigma$ poissonian error bars are overplotted for all bins that have a non-zero probability.  Smooth accretion tends to flatten galaxies over time.}
\label{fig:eigenval_smooth}
\end{figure*}

Focussing on more accurate morphological parameters, Fig.~\ref{fig:eigenval_smooth} displays the time evolution -- for galaxies which do not merge -- of the cumulative PDFs of the principal semi-axis ratios $\xi_{1}=c/a$,  $\xi_{2}=c/b$ and  $\xi_{3}=b/a$ with $a>b>c$ of the inner half-mass stellar component, derived by calculating the inertia tensor of the galaxy (see Appendix~\ref{sec:math}). One can see 
from this figure that, while $\xi_{3}$ tends to remain constant over cosmic time, with a value strongly peaked at 1 (large axis equals to intermediate axis), both average values of $\xi_{1}$ and $\xi_{2}$ decrease at an average rate of 
almost $10 \%$ per Gyr, from 0.64 and 0.74 down to 0.54 and 0.64 respectively in the 4 Gyr which separate $z=5.2$ and $z=1.2$. For reference, note that an infinitely thin and homogeneous disk has $\xi_{1}=\xi_{2}=0$ and $\xi_{3}=1$ while a perfect sphere has $\xi_{1}=\xi_{2}=\xi_{3}=1$. Our result indicates that smooth accretion and consecutive in situ star formation tend to flatten galaxies over time along the minor axis, which coincides with the spin axis. However, this accretion has no significant effect on the circularity of galaxy disks. 

\begin{figure*}
\center \includegraphics[width=2.05\columnwidth]{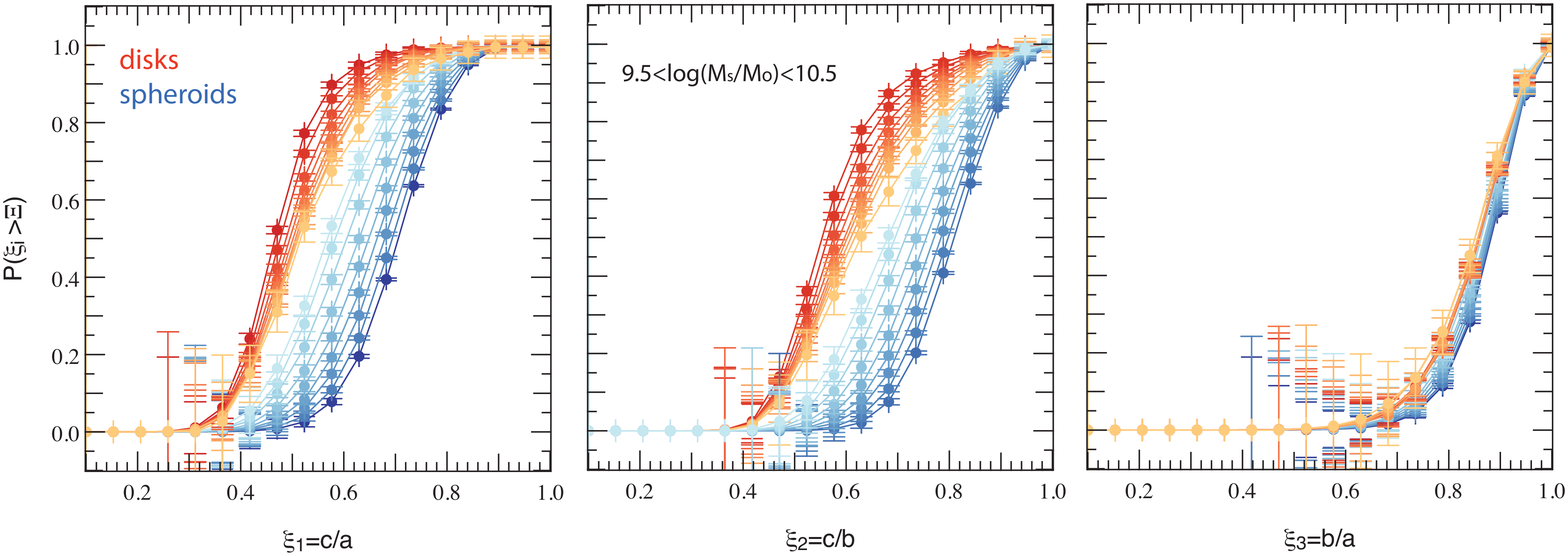}
\center \includegraphics[width=2.05\columnwidth]{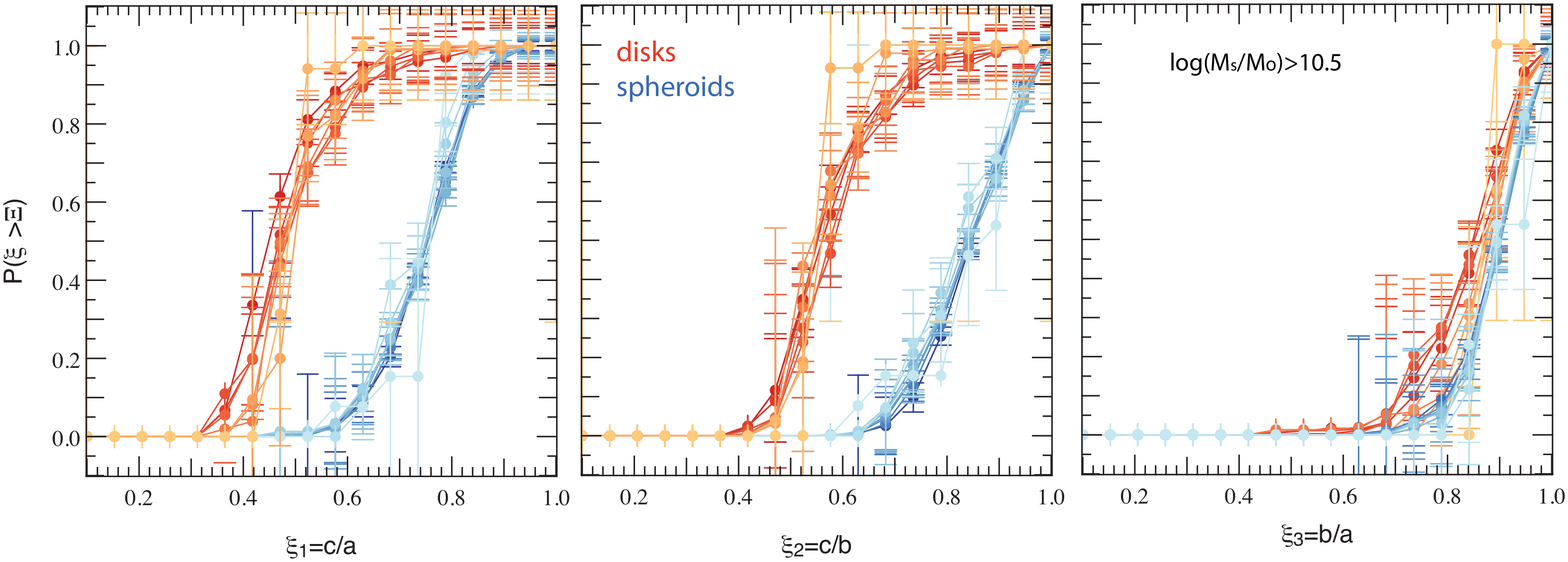}
  \caption{Cumulative PDFs of the principal semi-axis ratios $\xi_i$ for galaxies which do not merge. Colors evolve with cosmic time from dark red ($z=5.2$) to light orange ($z=1.2$) for disks and navy blue ($z=5.2$) to light blue ($z=1.2$) for spheroids, with an average time step of $\sim$ 200 Myr. The $1\sigma$ poissonian error bars are overplotted for all bins that have a non-zero probability.  \emph{Top panel}: galaxies with $9.5<\log(M_{s}/\rm M_\odot)<10.5$. \emph{Bottom panel}: galaxies with $\log(M_{s}/\rm M_\odot)>10.5$ (spin flip and cold/hot mode for accretion transition mass: see text for detail). Smooth accretion has little impact on the morphology of galaxies above the transition mass but clearly flattens galaxies below it.}
\label{fig:eigenval_smooth_tot}
\end{figure*}

This morphological transformation strongly depends on the mass and morphology of galaxies. As explained in Appendix~\ref{sec:math}, we define spheroids as galaxies with $\xi_{1}>0.7$ and $\xi_{2}>0.8$ and disks as galaxies with $\xi_{1}<0.45$ and $\xi_{2}<0.55$. Fig.~\ref{fig:eigenval_smooth_tot} displays the evolution of the principal semi-axis ratio PDFs for galaxies classified as spheroids (blue curves and symbols) and disks (red curves and symbols) pre-merger (galaxies are excluded from the sample when they merge), and for two different mass bins. The upper panels focus on galaxies with a stellar mass comprised between $10^{9.5} M_{\odot}$ and $10^{10.5} M_{\odot}$, the lower panel on more massive galaxies with $M_{s}>10^{10.5} \, \rm M_{\odot}$. This mass threshold corresponds to the transition mass above which galaxies embedded in filaments decouple from their vorticity quadrant and display a spin perpendicular to their closest filament~\citep[see][]{duboisetal14}, and also to the transition in gas accreted onto the galaxy between cold and hot mode \citep[e.g.][]{dekelbirnboim06,ocvirketal08}. The figure reveals that the decrease rate in $\xi_1$ and $\xi_2$ is much faster, around $20\%$ per Gyr for spheroids with masses below the transition mass (or from average values of $\xi_1 =0.7$ and $\xi_2 =0.8$  to 0.56 and 0.66 respectively). On the other hand, disks tend to thicken slightly on average (going from $\xi_1=0.46$ and $\xi_2=0.56$ to 0.5 and 0.64 respectively). This behaviour for the disks at least partially arises from the limited maximum spatial resolution of the simulation (1 kpc). By definition disks with scale heights below this value are artificially 'puffed up' to 1 kpc and any accretion of new material, no matter how dynamically cold, can only result in increasing this minimal numerical scale height. This especially alters the shape of small galaxies, for which the scale length is also poorly resolved. For galaxies with masses above the transition mass (bottom panels of Fig~\ref{fig:eigenval_smooth_tot}), we do not observe any significant impact of smooth processes on the morphology indicators which remain constant on average. 

Interpreting our results in the light of the scenario described by~\cite{codisetal12} and~\cite{laigle2015}, whereby small galaxies acquire their spin through angular momentum transfer from the vorticity quadrant they are embedded in, this "flattening" effect can be understood as the (re)-formation of disks in high vorticity regions at the heart of cosmic web filaments. 
In other words, smooth accretion tends to (re)-align galaxies with their nearest filament \citep{tillsonetal12, welkeretal14,danovichetal14,pichonetal2014} where the dominant component in this process, for galaxies below the transition mass, is coherent gas feeding from cold flows. At the opposite end of the mass spectrum, galaxies above the transition mass accrete material from multiple quadrants and/or smaller amounts of material along a unique filament. The angular momentum streamed to the core of these massive galaxies from multiple directions is more likely to cancel out, which results in little to no effect of smooth accretion on the morphology of the galaxy. These results reinforce earlier findings that the underlying cosmic web plays a major role in shaping galaxy properties.

\begin{figure*}
\includegraphics[width=0.65\columnwidth]{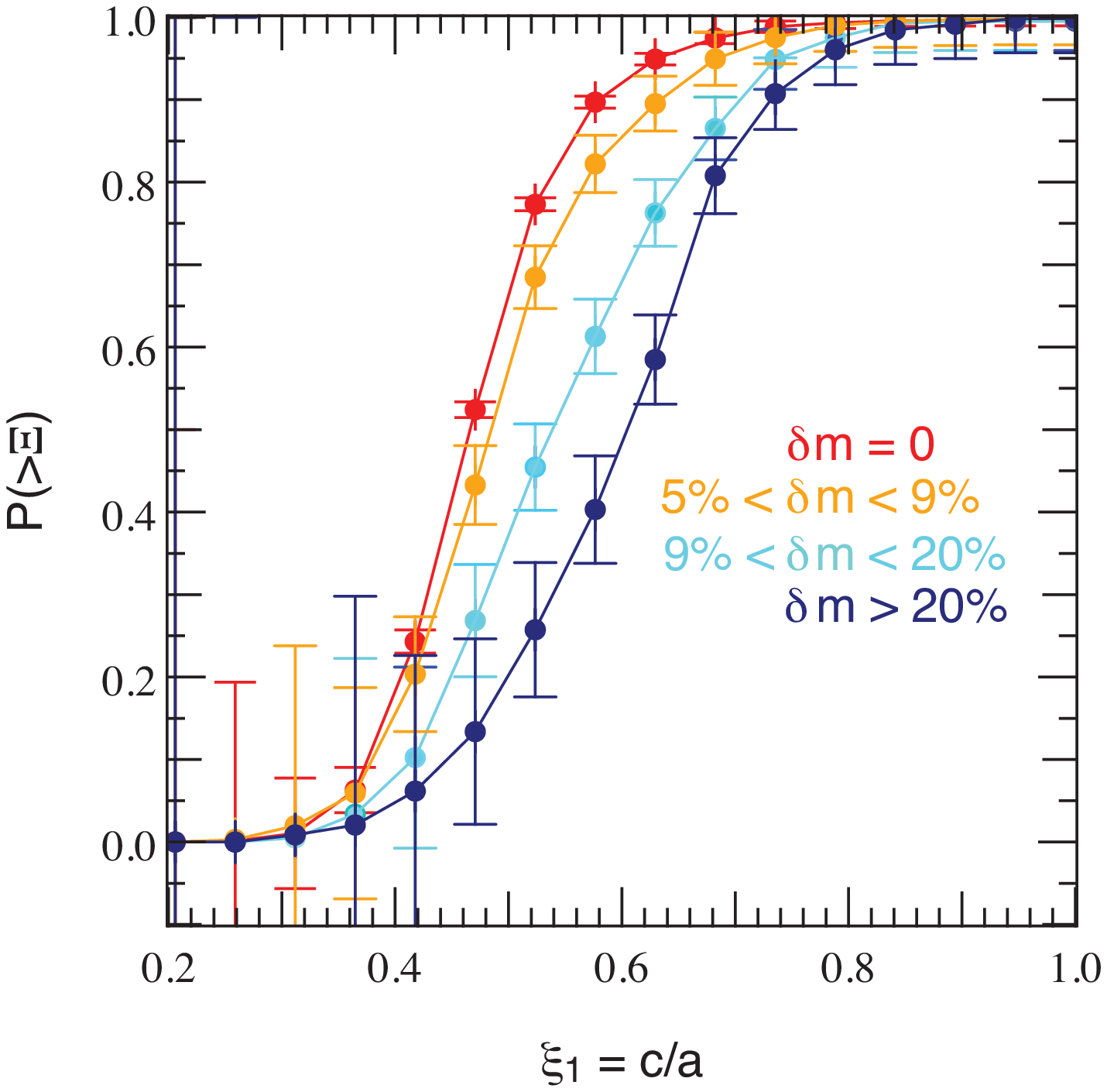}
 \includegraphics[width=0.65\columnwidth]{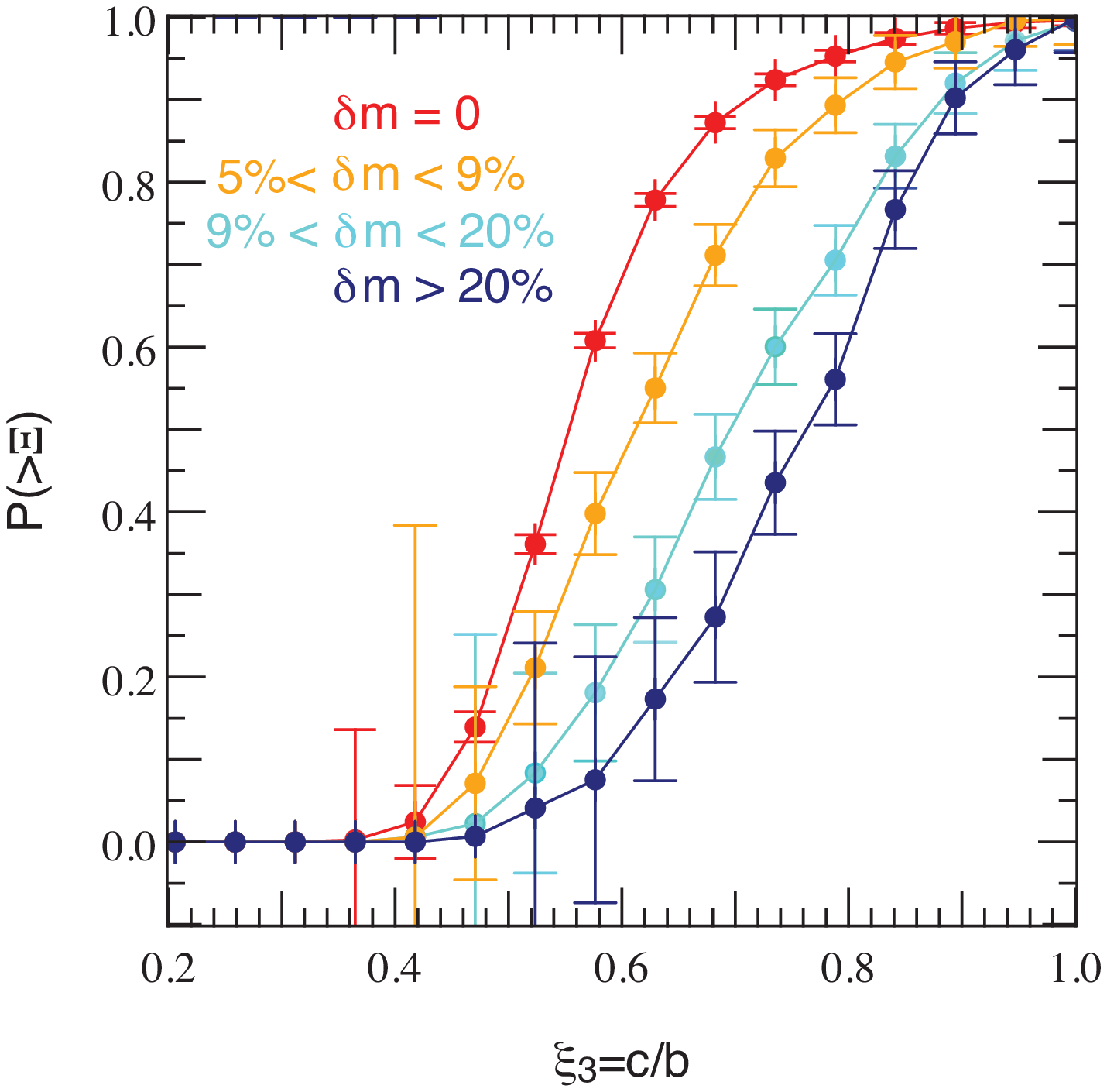}
  \includegraphics[width=0.65\columnwidth]{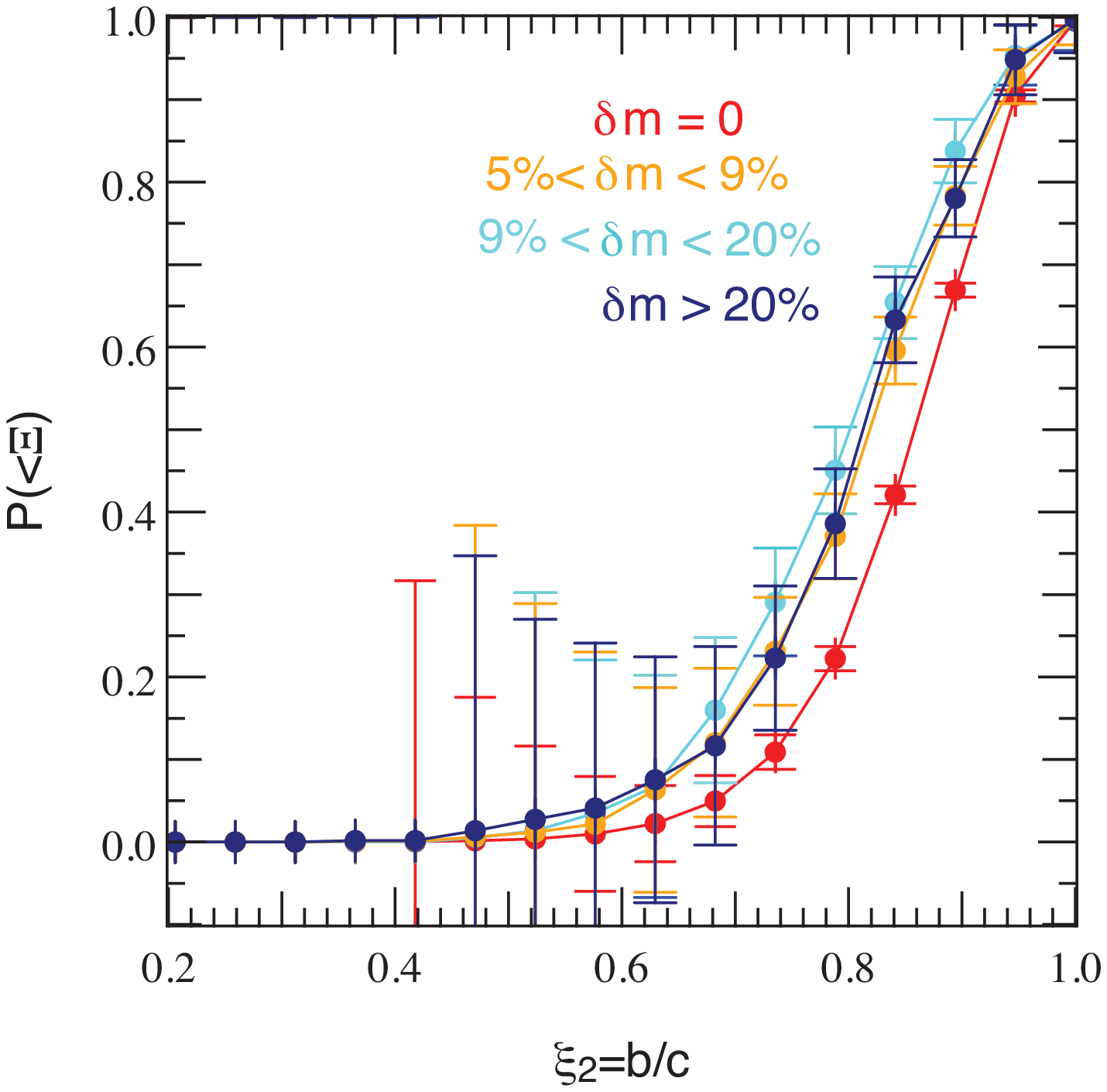}
  \caption{Cumulative PDFs of the morphology indicators $\xi_i$ for different merger mass ratios and for galaxies with $M_{s}>10^{10} \, \rm M_{\odot}$ identified as pre-merger disks. The poissonian $1\sigma$ error bars are indicated for all bins with more than 10 galaxies. The results are stacked for each time output between $z=5.2$ to $z=1.2$. Mergers broaden the morphology 
distribution and quantitatively destroy disks. This effect strengthens with increasing merger mass ratios.}
\label{fig:eigenval_spiral}
\end{figure*}

As can be seen on Fig.~\ref{fig:eigenval_spiral} (and \ref{fig:eigenval_ellipse}), mergers trigger very different evolutions for disk galaxies (respectively spheroids). This figure showcases a qualitative difference: contrary to smooth accretion, both minor and major mergers strongly change the galaxy morphology, leading to much more spheroidal/elliptical structures. However, this only occurs
for disks: as can be seen in Fig.~\ref{fig:eigenval_ellipse}, for galaxies initially identified as spheroids, minor mergers behave more like smooth accretion, flattening the galaxy whereas major mergers 
preserve, by and large, their morphology. Looking at Fig.~\ref{fig:eigenval_spiral}, there are quantitative differences between minor and major mergers. Whilst major mergers clearly destroy disks (the average value of the PDF of $\xi_1=c/a$ shifts from 0.45 to 0.62, red and dark blue curves respectively), minor mergers have a more limited effect ($\xi_1$ PDF average value shifted from 0.45 to 0.52 only). Finally, the effect of very minor mergers (light orange curve) is closer to a thickening of the disks than an actual destruction of them and an alteration of the galaxy morphology. It is important to notice that all mergers also trigger an increase in the scatter of the distribution of galactic disks morphology indicators $\xi_i$, as the slope of the PDFs becomes shallower for mergers than smooth accretion. This effect is stronger for higher merger mass ratios.

These findings corroborate the view that major \emph{and} multiple minor mergers can lead to galaxies with similar morphologies, destroying disks and turning them into spheroids~\citep[e.g.][]{bournaudjog07}. This allows to overcome the 
tension occasioned by the paucity of major mergers: minor mergers are much more frequent events, allowing for the formation of a much larger spheroid population. An illustration of this phenomenon 
is given in Fig.~\ref{fig:visual} which depicts rest-frame false color images of a sample of disk galaxies in the Horizon-AGN simulation before and after major/minor merging observed through the u, g and i filters.

\section{Conclusion}
\label{section:conclusion}

Galaxy growth in the cosmic web involves a wide range of processes from anisotropic accretion to supernovae and AGN feedback whose effect can
 either add up or cancel one another, resulting in the observed diversity in morphologies, kinematics and colors of galaxies. 
While the interplay between these phenomena is undoubtedly complex, the approach we implemented in this work which consists in focussing on a couple of well defined 
processes (smooth accretion against mergers) and identify their impact -- whether re-enforcing or competing -- on specific galactic properties (size, morphology) 
still yields some interesting results: 
\begin{itemize}
\item
Mergers and smooth accretion augment galaxy masses across the peak of cosmic star formation history, in amounts that are statistically comparable. As a result, at $z=1.2$, galaxies with $M_{s}>10^{10} \, \rm M_{\odot}$ have acquired $55 \%$ of their stellar mass via smooth accretion and $45\%$ via mergers. However, while smooth accretion is a steady process with regular impact on stellar mass over cosmic history, mergers are violent processes which occur on average twice in the history of a galaxy over this epoch. 
\item
Mergers and smooth accretion augment galaxy sizes across the peak of cosmic star formation history, especially major mergers, but this growth strongly depends on redshift and gas fraction. We also found that while mass is accreted, the mean density also rises for galaxies which are pre-merger disks, suggesting a gravitational contraction during the merger phase, while the inverse is true for pre-merger spheroids which on average expand after merging.
\item
For mergers of mass ratio $\delta m$, the relative increase in radius is found to evolve as a power law of the stellar mass $r_{0.5}=M_s^{0.65+\delta m}$ while smoothly accreted material of comparable mass ratio proves to be much less efficient in growing galaxy radii $r_{0.5}=M_s^{0.3\delta m}$. Moreover, while the growth of spheroid sizes shows little dependence on $\delta m$ ($r_{0.5} \propto  M_s$), -- even for smallest minor mergers which is consistent with the idea that material is then smoothly accreted within the galactic plane --, disks show a stronger dependence on $\delta m$,  even contracting trend when subjected to minor mergers. We interpret this result as the destruction of disks and redistribution of their stellar component in a more tightly packed spheroidal volume, which causes the effective half mass radius to decrease even though the amount of mass accreted actually increases.
\item
Gas fraction also plays an important part in determining the size growth consecutive to mass accretion. As expected, gas dominated mergers induce a much more limited growth in size than star dominated ones.
In such gas rich mergers, the remnant appears to be more compact. We interpret this as the result of gas shocking, losing angular momentum and being transported to the
central parts of the galaxy where it forms stars, seemingly triggering a gravitational contraction of the galaxy. On the opposite, star dominated mergers (with $f_{\rm gas}<0.2$) induce an increased growth in radius with no significant dependence on merger mass fraction, but the steepest dependence on stellar mass that we measure ($r_{0.5}=M_s^{2}$).

\item
These accretion processes are found to have a strong impact on galaxy morphologies. Smooth accretion tends to flatten small galaxies along their spin axis, consistent with the idea that those galaxies are embedded in a vorticity quadrant of cosmic filament which feeds them angular momentum coherently along the filament direction. This effect is even clearer for the sub-sample of spheroids fed by this smooth accretion which evolve to resemble the disk population in just over 2 Gyr. In contrast, mergers tend to destroy disks and form spheroids, except for very minor mergers -- which only thicken them --, in agreement with the idea that in this case the satellite is slowly stripped from its gas and stars in the galactic plane of the main progenitor. But our main result is that minor mergers are responsible for a comparable amount of disk destruction than major mergers, coupled with a strong contraction effect when the minor merger happens to a gas-dominated ($f_{\rm gas}>0.6$) galaxy. 

\item
These results altogether statistically favour a scenario whereby galaxies grow their stellar mass by smooth accretion of gas, in situ formation and mergers in comparable amounts, but grow in size mostly through merging: disk (gas-dominated) galaxies merge to become more compact spheroids while spheroids lose their compactness through these same minor mergers. Occasionally, dramatic growth in size through rare major mergers and multiple, gas-deprived minor mergers happens. Non-merging spheroids with masses up to a transition mass around $10^{10.5} \, \rm M_{\odot}$ then rebuild disks from coherent smooth accretion. Above this mass the coherence of streams is lost and morphology is preserved.
\end{itemize}

Though this first study supports -- in a full cosmological context using the \hagn\!\! simulation -- the consistent galaxy growth model that has emerged from previous numerical studies of different types of mergers, further investigation is required to extend these results down to $z=0$ and to specify in detail the role played by galactic physics -- more specifically supernovae and AGN feedback -- in shaping these results. Analyzing more specific merger parameters such as the impact parameter and the orbital-to-intrinsic angular momentum transfer rate will also be necessary to understand the scattering of the morphology and size distributions induced by mergers and understand their overall impact on observed galaxies in the local Universe. Finally, the internal kinematics of the galaxy population also need to be examined more closely.

\vskip 0.1cm
\section*{Acknowledgments}
{\sl This work was granted access to the HPC resources of CINES (Jade) under the allocation 2013047012 and c2014047012 made by GENCI.
This research is part of the Spin(e) (ANR-13-BS05-0005, \url{http://cosmicorigin.org}) and Horizon-UK projects. 
Let us thank D.~Munro for freely distributing his {\sc \small  Yorick} programming language and opengl interface (available at \url{http://yorick.sourceforge.net/}). 
We  warmly thank S.~Rouberol for running  the {\tt Horizon} cluster on which the simulation was  post-processed.  Part of the analysis was also performed using the DiRAC facility jointly funded by BIS and STFC.
The research of JD is supported by Adrian Beecroft, The Oxford Martin School and STFC. 
CP thanks Churchill college for hospitality while this work was finalized.
}
\vspace{-0.5cm}

\bibliographystyle{mn2e}
\bibliography{author}

\appendix

\section{Inertia tensor}
\label{sec:math} 

    The inertia tensor $I_{ij}$ of a galaxy is computed from its star particle distribution (indexed by $l$) and calculated at the centre of mass of the galaxy, according to the definition: $I_{ij} = \Sigma_{l} m^{l}(\delta_{ij}.(x^{l}_{k}.x^{l}_{k})-x^{l}_{i}.x^{l}_{j})$, where $m_{l}$ is the mass of star particle $l$ and $x^{l}_i$ its position in the barycentric coordinate system of the galaxy. As a 3x3 real symmetric matrix, the inertia tensor can be diagonalized, with its eigenvalues $\lambda_{1} > \lambda_{2} > \lambda_{3}$ being the moments of inertia relative to the basis of principal axes $e_{1}$, $e_{2}$ and $e_{3}$.
The lengths of the semi-principal axes $a$, $b$ and $c$ (with $a>b>c$) are straightforwardly derived from the moments of inertia: 
\begin{eqnarray}
a = (5/M_{0.5})\sqrt{\lambda_{1}+\lambda_{2}-\lambda_{3}} \, , \: \rm along \: e_{3} \nonumber \, , \\
b = (5/M_{0.5})\sqrt{\lambda_{1}+\lambda_{3}-\lambda_{2}} \, , \: \rm along \: e_{2} \nonumber \, , \\
c = (5/M_{0.5})\sqrt{\lambda_{3}+\lambda_{2}-\lambda_{1}} \, , \: \rm along \: e_{1} \nonumber \, .
\end{eqnarray}


Therefore the morphology of a galaxy can be easily interpreted using its axis ratios: $\xi_1= c/a$, $\xi_2= c/b$ and $\xi_3= b/a$. A perfectly round and infinitely thin disk has $\xi_1= 0$, $\xi_2= 0$  and $\xi_3= 1$.  For a Milky Way like galaxy (including stars from the inner bulge+thin disk+bar), one gets  $\xi_1= 0.06$, $\xi_2= 0.07$ and $\xi_3= 0.98$.
The limited spatial resolution (1 kpc) of the Horizon-AGN simulation, prevents us from obtaining disks as thin as these. We therefore identify disks with our most flattened ellipsoids.
More specifically,  we adopt $\xi_{1}<0.45$ and $\xi_{2}<0.55$ as a definition for disks and $\xi_{1}>0.7$ and $\xi_{2}>0.8$ to define spheroids. Other galaxies are simply classified as 
ellipsoids. Though these cuts may appear a crude approximation, they are actually quite consistent with 3-D axis ratios reconstructed from observations~\citep{Lambasetal92}. 
Note that we also define the morphology of our galaxies using star particles enclosed within their half mass radius sphere. We found that this is more robust than using all the star particles 
identified by our halo finder especially for post-merger remnants, as these can exhibit elongated tidal features which persist for a considerable amount of time.

\section{Gas content}
\label{sec:gas} 

\begin{figure}
\center \includegraphics[width=1.1\columnwidth]{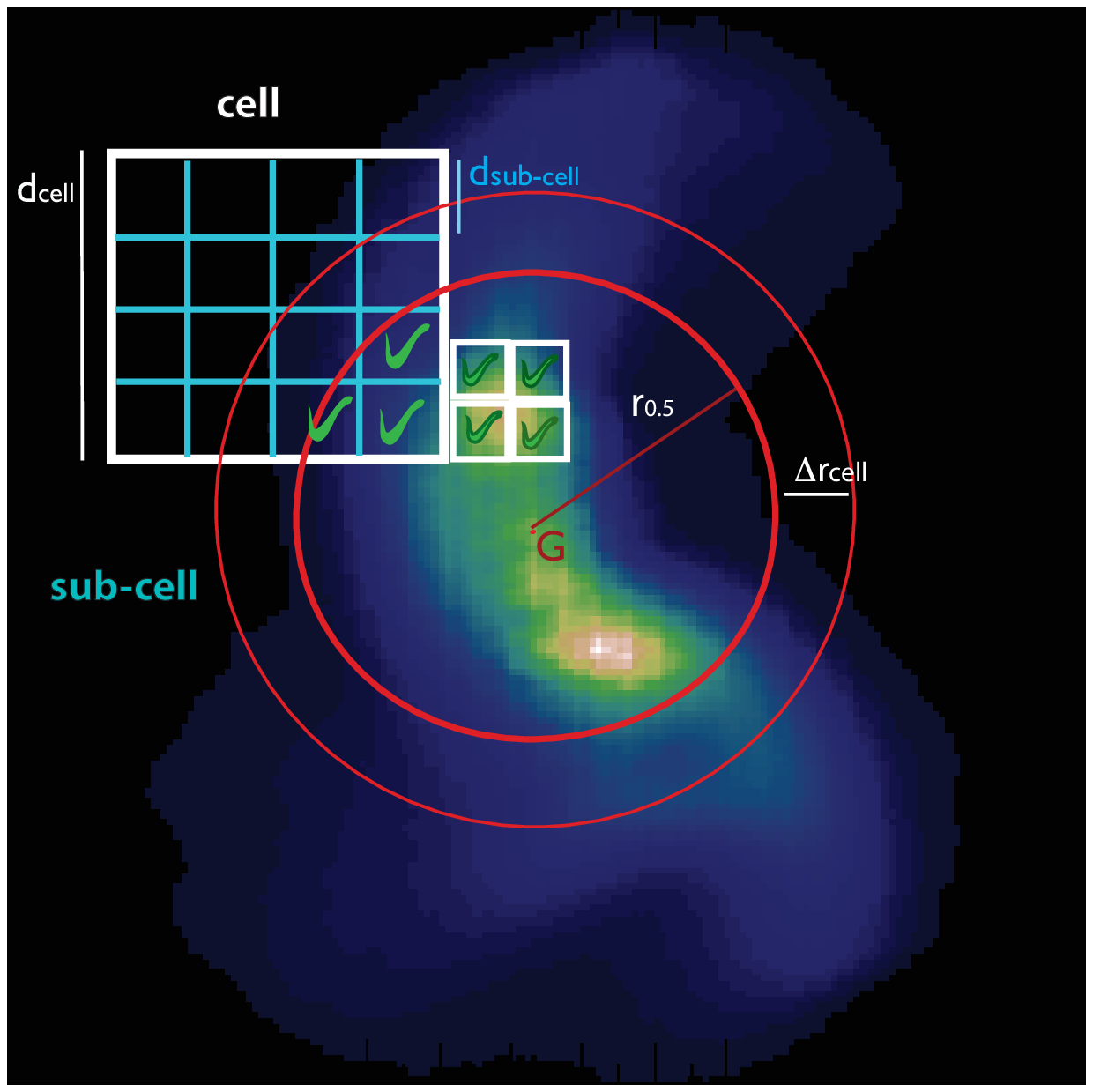}
  \caption{2D sketch of the gas cell assignment procedure for one galaxy from Horizon-AGN (shown as a face-on projected gas density map). The thick red circle represents the effective radius $r_{0.5}$ around the galactic center of mass and the white squares the AMR grid with different levels of refinement. The green tick indicates when a cell or sub-cell is counted as belonging to the galaxy (see text for detail).}
\label{fig:extraction}
\end{figure}

Fig.~\ref{fig:extraction} displays a sketch of the systematic procedure used to extract gas content from AMR cells for all galaxies in Horizon-AGN.

\begin{figure}
\center \includegraphics[width=0.85\columnwidth]{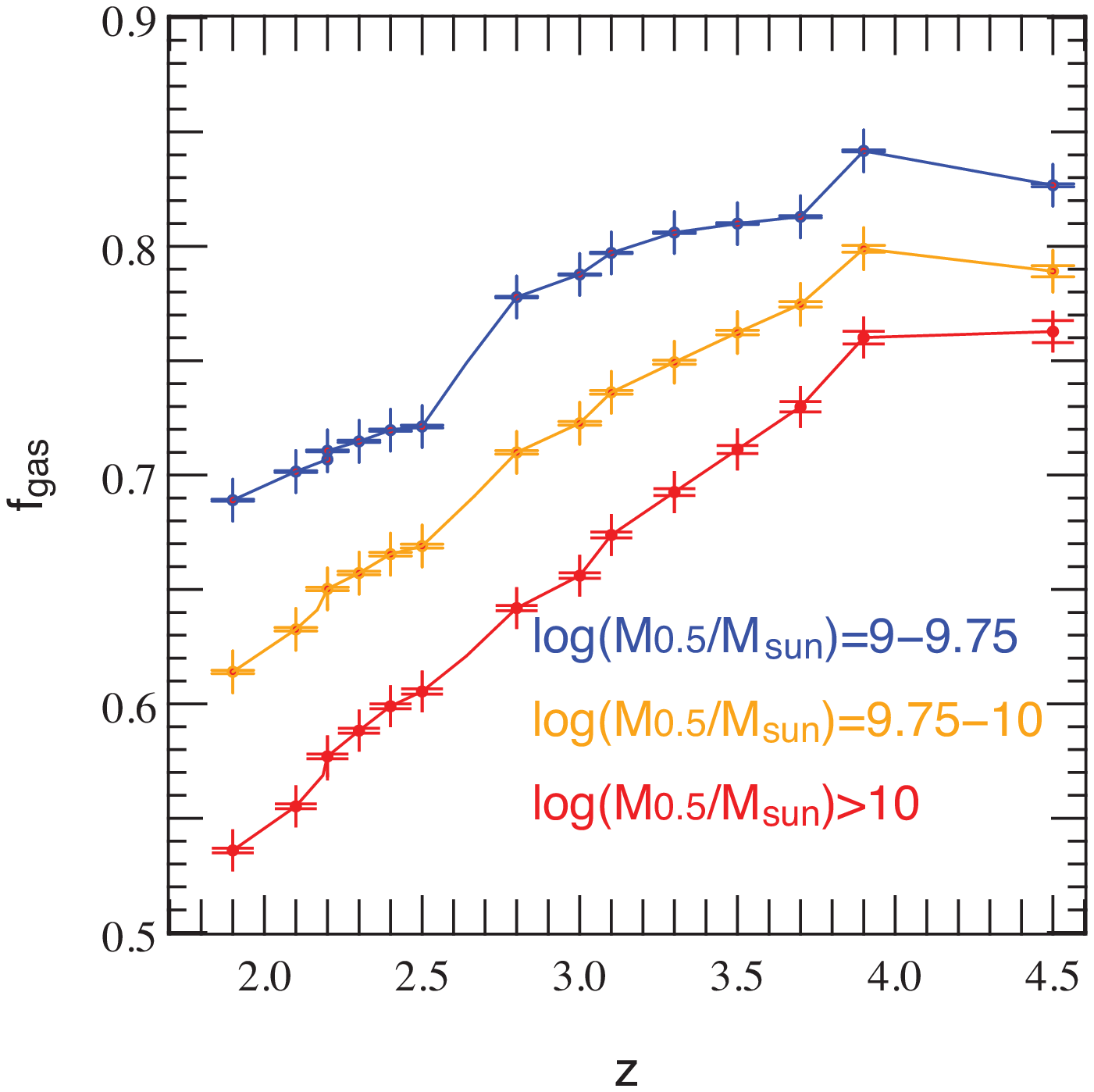}
  \caption{Evolution of the gas fraction $f_{\rm gas}$ in the redshift range $1.8<z<4.5$ for different mass bins, where $M_{0.5}$ is the stellar mass enclosed within the half mass sphere 
 and ${\rm M_{sun}}$ the mass of the sun. Results are consistent with previous simulations \citep[e.g.][]{popping14}. $f_{\rm gas}$ decreases with redshift as star formation consumes the available gas and/or feedback blows it out of the galaxies.}
\label{fig:gas_fraction}
\end{figure}

Fig.~\ref{fig:gas_fraction} shows the evolution of the average gas fraction across the peak of cosmic star formation history for galaxies of different masses. Our results are consistent with previous numerical investigations \citep[e.g.][]{popping14}. $f_{\rm gas}$ decreases with redshift as star formation consumes and feedback expels the available gas. 

For each galaxy in the sample, we define the maximum radius $r_{\rm max}$ as the distance between the galactic center of mass (COM) and the furthest star particle, the effective radius $r_{0.5}$ as the half stellar mass radius and  $\Delta r_{\rm cell} = r_{0.5}/10$. AMR cells with a size $d_{\rm cell}$ larger than $\Delta r_{\rm cell}$ are subdivided in $2^{3n_c}$ sub-cells with $n_c$ such that $d_{\rm sub-cell} < \Delta r_{\rm cell}$. AMR cells counted as belonging to the galaxy are: 1) AMR cells with a size $ d_{\rm cell} < \Delta r_{\rm cell} $ and a center within the sphere of radius $r_{0.5}$ centred on the COM 2) sub-cells of larger AMR cells with a length $ d_{\rm cell} < \Delta r_{\rm cell} $ and a center within the sphere of radius $r_{0.5}$ centred on the galaxy COM. This procedure is illustrated on  Fig.~\ref{fig:extraction} which shows a 2D sketch of the cell selection process on a face on projected gas density map for a galaxy from Horizon-AGN with a post-merger sub-structure at $z=3$.

\section{Mass}
\label{sec:mass}

\begin{figure}
\includegraphics[width=0.85\columnwidth]{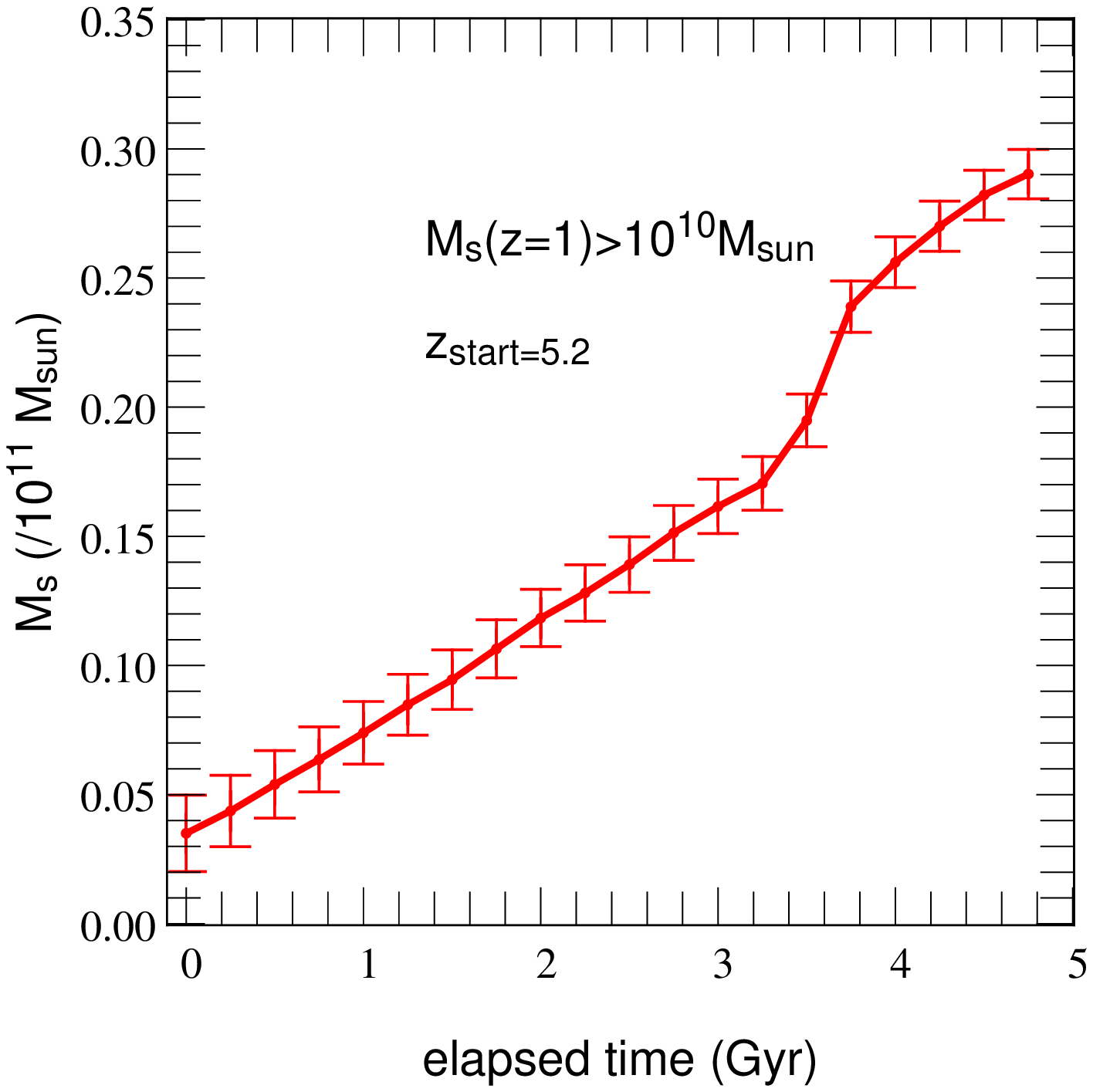}
  \caption{ Evolution of the average stellar mass of galaxies with $M_{s} > 10^{10} \, {\rm M_\odot}$ at $z=1.2$. }
\label{fig:mass}
\end{figure}

Fig.~\ref{fig:mass} shows the evolution of the average stellar mass of galaxies with $M_{s} > 10^{10} \, {\rm M_\odot} $ at $z=1.2$, across the peak of cosmic star formation 
history. Note that the stellar mass growth of galaxies remains steady for most of the evolution. The knee at redshift $z=1.5$ corresponds to a peak in the merger rate and smooth accretion observed at the same redshift (see Fig.~\ref{fig:rates_smooth}), which is due to the extra level of refinement added at this particular redshift.

\section{Morphology}
\label{sec:morphall} 

\begin{figure}
\center \includegraphics[width=0.85\columnwidth]{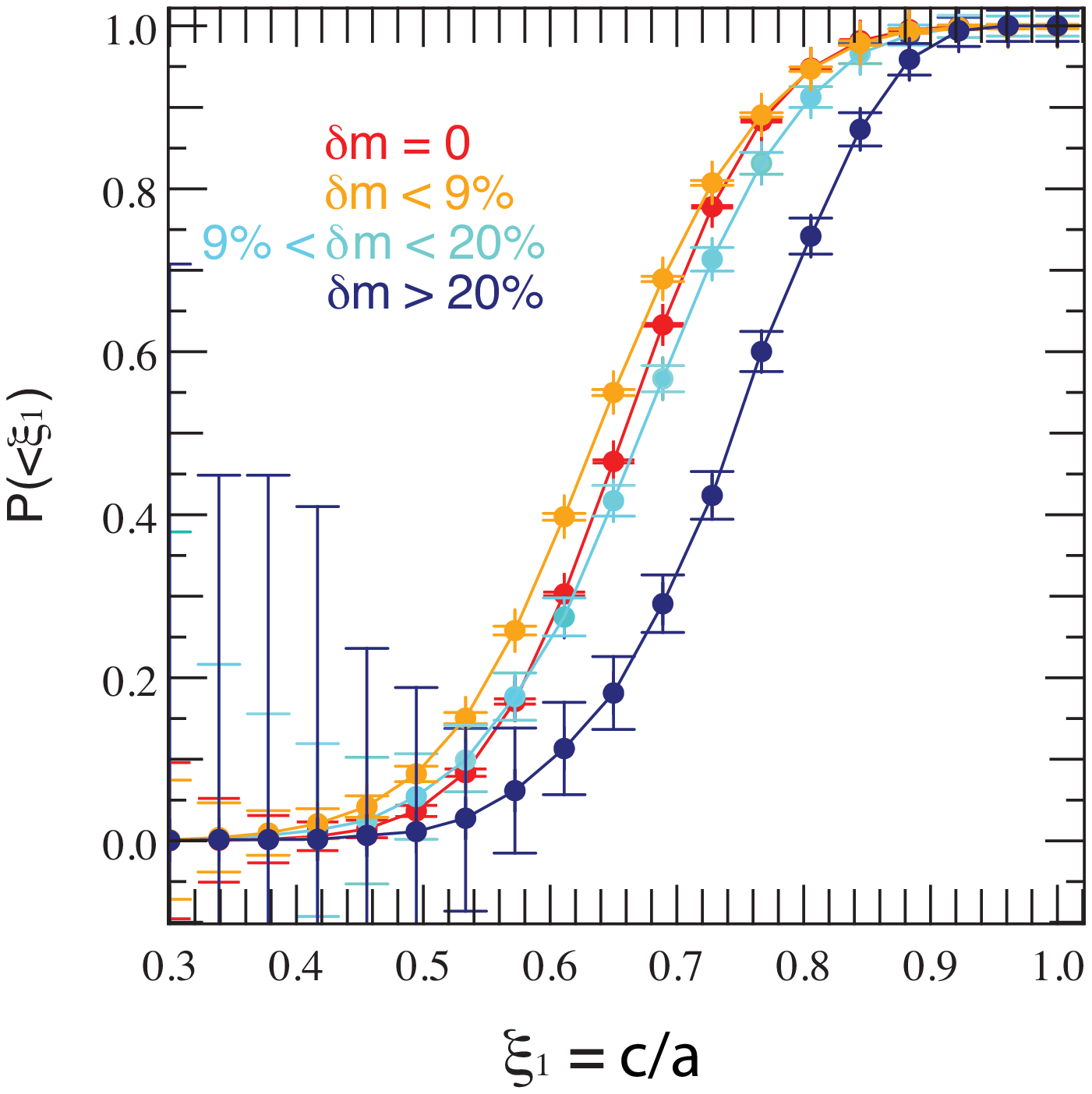}
  \caption{Cumulative PDF of $\xi_1=c/a$ over one time step, with all 
timesteps between $z=5.2$ and $z=1.2$ stacked, for different merger mass ratios and for galaxies with $M_{s}>10^{10} M_{\odot}$.}
\label{fig:eigenval_mergers}
\end{figure}

\begin{figure}
\center \includegraphics[width=0.8\columnwidth]{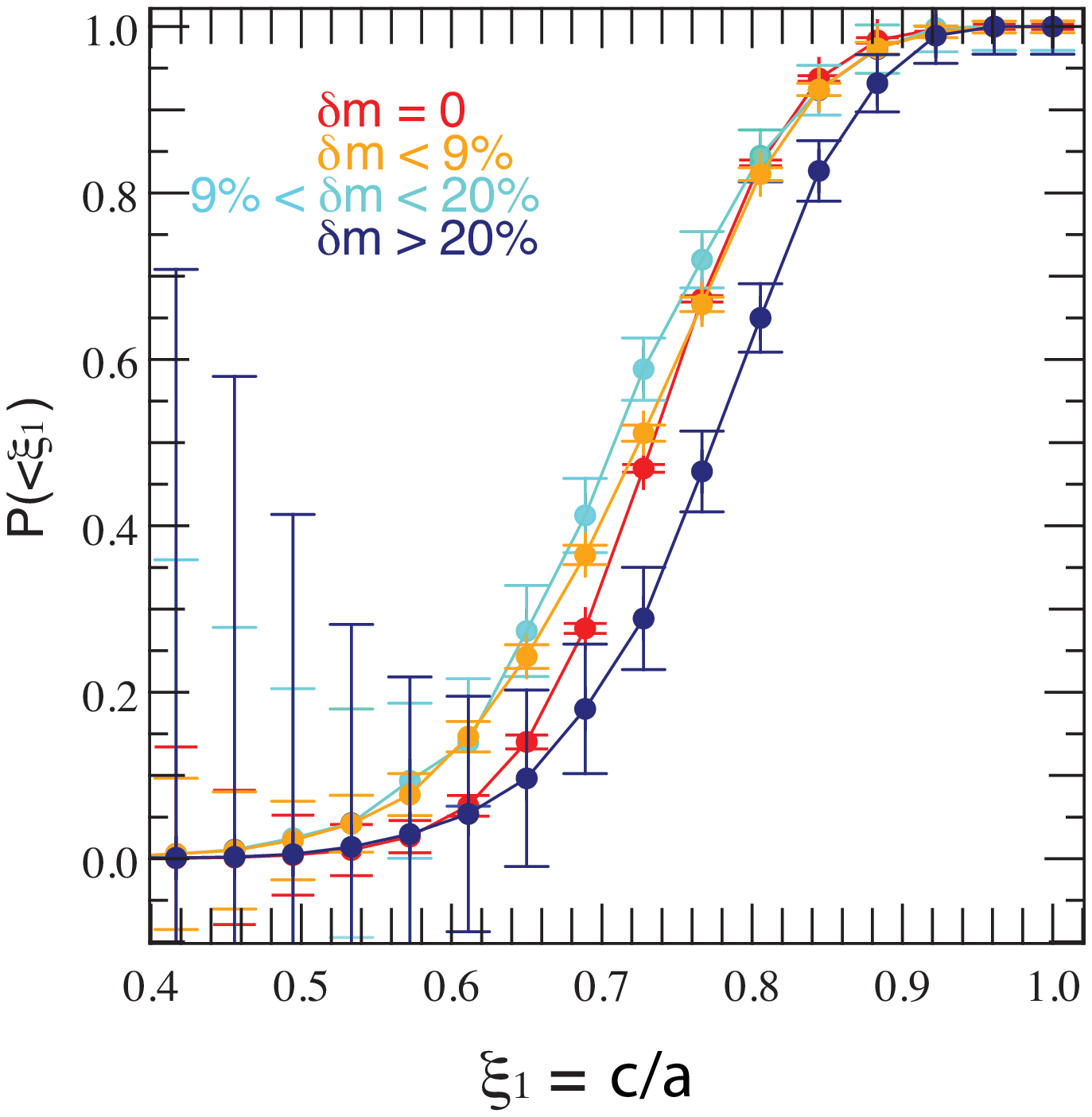}
  \caption{Same as Fig.~\ref{fig:eigenval_mergers} but for galaxies which are classified as spheroids before they merge (see text for detail).}
\label{fig:eigenval_ellipse}
\end{figure}

Fig.~\ref{fig:eigenval_mergers} shows the cumulative probability distributions of the morphologic ratio $\xi_1=c/a$ over one time step, with all 
timesteps between $z=5.2$ and $z=1.2$ stacked, for different merger mass ratios. This is plotted for all galaxies 
regardless of their morphology. Fig.~\ref{fig:eigenval_ellipse} displays the same quantity for galaxies identified as spheroids, 
i.e. with morphologic ratios $\xi_{1}>0.7$ and $\xi_{2}>0.8$, before they merge.

\end{document}